\documentclass[letterpaper,twocolumn,10pt]{article}
\usepackage{usenix2019_v3}

\usepackage{tikz}
\usepackage{xfrac}
\usepackage{amsmath}
\usepackage{microtype}
\usepackage{graphicx}
\usepackage{booktabs} %
\usepackage{algorithm}
\usepackage{algorithmic}
\renewcommand{\algorithmiccomment}[1]{\textit{#1}}
\usepackage{mathtools}
\usepackage[textsize=footnotesize, linecolor=magenta, bordercolor=magenta, backgroundcolor=white, textwidth=35pt]{todonotes}
\usepackage{amssymb,amsmath,booktabs}
\usepackage{enumitem}
\usepackage{titlesec}
\titlespacing*{\paragraph}{0pt}{0.5ex plus 0.2ex minus 0.2ex}{0.5ex}

\usepackage{authblk}  %

\newcommand{\brown}[1]{#1}
\newcommand{\green}[1]{#1}
\definecolor{darkred}{RGB}{205, 92, 92}
\definecolor{lightred}{RGB}{231, 143, 137}
\definecolor{forestgreen}{RGB}{160, 218, 184}
\definecolor{lightblue}{RGB}{114, 159, 220}
\newcommand{\forward}[1]{\textcolor{lightblue}{#1}} 
\newcommand{\backward}[1]{\textcolor{darkred}{#1}}
\newcommand{\dgrad}[1]{\textcolor{lightred}{#1}}
\newcommand{\wgrad}[1]{\textcolor{forestgreen}{#1}}

\usepackage{array}
\newcolumntype{A}{>{\small}c}
\newcolumntype{B}{>{\footnotesize}l}

\usepackage{filecontents}

\begin{document}

\date{}

\title{\Large \bf CrossPipe: Towards Optimal Pipeline Schedules for Cross-Datacenter Training}

\author[1]{Tiancheng Chen}
\author[1]{Ales Kubicek}
\author[1]{Langwen Huang}
\author[1]{Torsten Hoefler}

\affil[1]{Department of Computer Science, ETH Zürich}

\renewcommand\Authands{ and }  %

\maketitle

\begin{abstract}

Training large language models (LLMs) now requires resources that exceed a single datacenter, making cross-datacenter strategies increasingly crucial. We present CrossPipe, a framework designed to optimize model training across geographically distributed datacenters by explicitly modeling and mitigating the impact of network latency and limited bandwidth. It enables unified analysis and optimization incorporating both pipeline parallelism (PP) and opportunities for overlapping data parallelism (DP) communication. CrossPipe generates optimized pipeline schedules using either solver-based optimal or fast near-optimal greedy algorithms, built upon a flexible execution engine that separates scheduling logic from communication details. Our evaluation shows that CrossPipe reduces training time by up to 33.6\% compared to traditional pipeline schedules under identical memory constraints. When memory constraints are relaxed, CrossPipe maintains strong performance despite communication delays, approaching the efficiency of idealized schedules without delays. CrossPipe offers improved scalability and resource utilization, particularly in environments with high network latency or limited bandwidth.

\end{abstract}

\vspace{10pt}

\section{Introduction}
\label{sec:introduction}
Large language models (LLMs) have revolutionized natural language processing, demonstrating remarkable capabilities in tasks such as text generation, translation, and question answering. These models, trained on massive datasets, exhibit sophisticated context understanding and generate human-like responses. Their applications span scientific research~\cite{thirunavukarasu2023large}, content generation~\cite{li2024pre}, and personal assistants~\cite{li2024personal}. LLM training consists of three stages: pre-training, fine-tuning, and alignment, with pre-training consuming the majority of computational resources~\cite{duan2024efficient,han2024parameter,wang2023aligning}.

As LLM performance scales with model size and data volume, computational demands have increased exponentially. Recent studies estimate that training compute for state-of-the-art models quadruples annually~\cite{epoch2024trainingcomputeoffrontieraimodelsgrowsby45xperyear}, necessitating both vertical scaling (faster accelerators) and horizontal scaling (distributed computation). 
While GPU performance and energy efficiency continue to improve steadily~\cite{epoch2023trendsinmachinelearninghardware}, the power and infrastructure required to support LLM training is growing even faster. If current trends hold, GPU counts must nearly triple each year, with power consumption rising accordingly.

The escalating computational demands of LLMs are straining existing infrastructure, particularly power supply systems. To address these energy requirements, companies like Microsoft, Google, and Amazon are turning to nuclear energy sources to power their new AI datacenters~\cite{reuters2024microsoft, daSilva2024google, Moseman2024}, emphasizing the need for reliable and high-capacity power sources.
\brown{Scaling a single datacenter introduces challenges including local power limitations and increased vulnerability to outages. 
Report suggests that deploying multiple smaller facilities is more practical than scaling a single massive one~\cite{bloomberg2024}.}

As LLM scale grows, multi-datacenter training is becoming essential~\cite{albergotti2024microsoft}, distributing both compute and energy loads. 
However, geographic distribution introduces significant communication inefficiencies that must be addressed to support this shift. In the context of cloud-based training, allocating large blocks of GPUs in one region is often infeasible~\cite{10.1145/3642970.3655843}, making cross-regional GPU acquisition a practical alternative. The high cross-region communication cost poses challenges to the efficiency of existing training methods. \green{This work attempts to assess the impact of network inefficiencies to synchronous training tasks and improve the performance to reduce the cost and energy consumption.}

This paper introduces CrossPipe\footnote{The code is available at \url{https://github.com/spcl/crosspipe}.}, a framework that improves the efficiency of cross-datacenter (cross-DC) LLM training through the following contributions:

\begin{itemize}[noitemsep, topsep=0pt,leftmargin=*]
    \item \textbf{Analysis:} We present a comprehensive analysis of cross-DC training methodologies and show that pipeline parallelism is the most feasible approach in this setting.
    \item \textbf{Performance Model and Algorithm:} We present a latency and bandwidth-aware performance model specifically designed for the cross-DC environment. This model enables the co-optimization of pipeline schedules with potential data parallelism (DP) communication overlap, unifying the modeling of cross-DC PP and cross-DC DP. Next, we introduce a system-aware pipeline schedule generation algorithm: CrossPipe. The algorithm leverages either constraint optimization techniques to generate \textit{optimal} cross-DC pipeline schedules (Section~\ref{sec:cpoform}) or fast greedy algorithm to generate efficient and \textit{near-optimal} schedules (Section~\ref{sec:greedy_schedule}). 
    \item \textbf{Framework:} \brown{Finally, we propose and implement a flexible and easily extensible pipeline execution engine featuring a two-layer abstraction that decouples block scheduling from communication arrangement (detailed in Section~\ref{sec:exeplan}). This design enables efficient deployment of different pipeline schedules, including those generated by CrossPipe. }
\end{itemize}

\section{Cross-DC Training}

\subsection{Parallelism Strategies} \label{sec:parallel_strategies}

Distributed LLM pre-training~\cite{bennun2018demystifyingparalleldistributeddeep} employs a combination of different parallelism strategies (termed \textit{hybrid parallelism}) to partition the workload across GPU clusters.  
Table~\ref{tab:symbols} lists symbols and notations used in this paper.

\textbf{Tensor Parallelism (TP)}~\footnote{Also known as Operator Parallelism.}: Splits each model layer across multiple GPUs~\cite{shoeybi2020megatronlmtrainingmultibillionparameter,korthikanti2022reducingactivationrecomputationlarge}, requiring extensive collective communication (e.g., Reduce-Scatter and Allgather~\cite{mpi31}) during both forward and backward passes. Due to limited opportunities for overlap~\cite{10.1145/3567955.3567959} and high communication costs, TP is typically restricted to high-bandwidth domains (e.g., NVLink~\cite{nvlink1}), making it unsuitable for spanning geo-distributed DCs.

\textbf{Pipeline Parallelism (PP)}: Divides the model layers into $n_{PP}$ \textit{stages}, with each stage assigned to a different GPU. Communication occurs only at stage boundaries via point-to-point send/receive of activations and gradients.

\textbf{Data Parallelism (DP)}: Replicates the full model on each GPU, where distinct batches are processed independently and gradients are synchronized across replicas. DP is usually applied with ZeRO~\cite{Rajbhandari_2020} to reduce memory redundancy. This work assumes DP with ZeRO stage 1, partitioning optimizer states without increasing communication overhead compared to vanilla DP. Higher ZeRO stages introduce extra communication with diminishing memory savings.

\textbf{Sequence Parallelism (SP)}~\footnote{Also known as Context Parallelism.}: Scales sequence dimension~\cite{liu2023ringattentionblockwisetransformers,jacobs2023deepspeedulyssesoptimizationsenabling} and is typically applied at the end of pre-training to increase model context window~\cite{dubey2024llama3herdmodels}.

\textbf{Expert Parallelism (EP)}: Distributes the expert MLPs in Mixture of Experts (MoE)~\cite{jacobs1991adaptive, shazeer2017outrageouslylargeneuralnetworks} models. The per-layer, high-volume, and dynamic Alltoall communication in EP makes it challenging to deploy on cross-DC links.

\textbf{Key Insight}: TP, SP, and EP introduce layer-wise communication with high frequency and/or volume, making them highly sensitive to the latency and limited bandwidth typical of cross-DC links. Therefore, PP and DP emerge as the primary candidates for cross-DC traffic, due to their less frequent (PP, DP) or point-to-point (PP) communication patterns.

\begin{table}[t]

\centering
\setlength{\tabcolsep}{1.5pt}
\footnotesize
\sf
\begin{tabular}{AB}
\toprule
\textbf{Notation} & \textbf{Description} \\
\midrule

$n_{DC}$ & \# of datacenters \\
\vspace{4pt}
$n_{\{TP,PP,DP\}}$ & TP, PP, DP size \\

$T_{\{F,B,W\}}$ & Runtime of F, B, W block \\
$M_{\{F,B,W,L\}}$ & Net memory change in F, B, W block and memory budget \\
$M_{L}$ & Memory limit per device \\
\vspace{4pt}

$\alpha$, $\beta$ & Communication latency and  inverse of bandwidth\\
$T_{\alpha},\; T_{\beta}$ & Communication cost matrix \\

$b,\; \hat{B}$ & Microbatch size and global batch size \\
$n_{mb},\; \epsilon$ & Number of microbatches per DP rank, and ratio $n_{mb}/n_{PP}$ \\

$s,\; d$ & Model sequence length and hidden dimension \\

$n_{sub}$ & Number of parts in a sub-block schedule (Section~\ref{sec:greedy_schedule}) \\

\bottomrule
\end{tabular}
\rm
\caption{
List of symbols and notations. 
}
\label{tab:symbols}
\end{table}

\subsection{Communication Model}\label{sec:comm_model}

\begin{figure}[!h]
\centering
\includegraphics[width=0.9\linewidth]{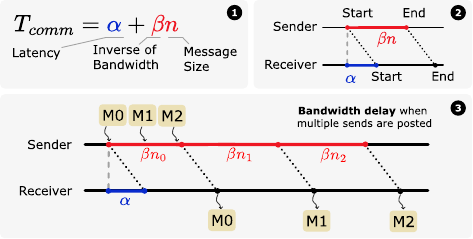}
\caption{
Alpha-Beta communication model \includegraphics[scale=1.1,trim=0 1 0 0]{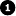} sending a single message \includegraphics[scale=1.1,trim=0 1 0 0]{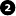}. When multiple pending messages are to be sent, the latter messages need to wait for the previous ones to be placed on the transmission link first. This results in an extra bandwidth delay \includegraphics[scale=1.1,trim=0 1 0 0]{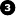}.
}
\label{fig:comm_model}
\end{figure}

In this work, we assume a small number of DCs (e.g., $n_{DC}\leq 4$). For modeling the communication time $T_{comm}$, we adopt the Alpha-Beta model, accounting for both latency ($\alpha$) and bandwidth limitations ($\beta$).  Multiple concurrent messages incur additional queuing delays, illustrated in Figure~\ref{fig:comm_model}.

\subsection{Distributed Training Infrastructure}\label{sec:infra}
\begin{figure}[!h]
\centering
\includegraphics[width=1.0\linewidth]{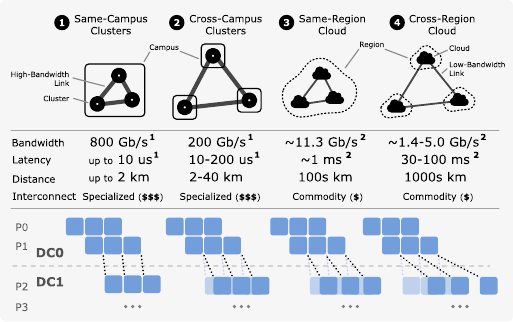}
\caption{
Cross-DC infrastructure setup types and their impact on the PP communication at DC boundaries. 
}
\label{fig:crossdclinks}
\end{figure}

Cross-DC infrastructure setups can be categorized into four primary types, as shown in Figure~\ref{fig:crossdclinks}. We differentiate between high-performance clusters and public cloud environments, further classified by geographic proximity: either nearby (same-campus, same-region) or distant (cross-campus, cross-region).

\begin{itemize}[noitemsep, topsep=0pt, leftmargin=*]
\item\textbf{Same-Campus Clusters} \includegraphics[scale=1.1,trim=0 1 0 0]{figures/icons/Circle1.pdf} setup represents tightly interconnected DCs on the same campus, typically connected via frontend network or additional switch layer. This setup features low latency~(up to 10 us~\footnote{\label{footnote:theory}Calculation based on the best commercially available hardware (NVIDIA LinkX / MetroX), assuming 5 ns/m.}) and high bandwidth~(800 Gb/s~\textsuperscript{\ref{footnote:theory}} per port), making communication overhead negligible, but still slightly higher than a single cluster. 

\item\textbf{Cross-Campus Clusters} \includegraphics[scale=1.1,trim=0 1 0 0]{figures/icons/Circle2.pdf} setup represents geographically distributed DCs (typically up to 40 km apart, using public products like NVIDIA MetroX) interconnected with high-bandwidth links~(200 Gb/s~\textsuperscript{\ref{footnote:theory}} per port). Network latency is bounded by the physical distance~(10-200 us~\textsuperscript{\ref{footnote:theory}}). 

\item\textbf{Same-Region Cloud} \includegraphics[scale=1.1,trim=0 1 0 0]{figures/icons/Circle3.pdf} setup represents closely allocated instances within the same cloud region. This setup features low bandwidth~(around 11.3 Gb/s~\textsuperscript{\ref{footnote:measured}})~\cite{strati2024ml} and higher latency (around 1~ms~\footnote{\label{footnote:measured}Measured.})~\cite{strati2024ml} compared to the setups discussed above due to the usage of less specialized networking hardware. 

\item\textbf{Cross-Region Cloud} \includegraphics[scale=1.1,trim=0 1 0 0]{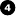} setup represents instances allocated across cloud regions or even continents. This setup inherits the same properties as a same-region cloud but with significantly higher latency~(30-100~ms~\textsuperscript{\ref{footnote:measured}})~\cite{strati2024ml} and even lower bandwidth~(1.4-5.0~Gb/s~\textsuperscript{\ref{footnote:measured}})~\cite{strati2024ml} due to distance. 
\end{itemize}

\begin{table}[h!]
\centering
\caption{\green{The hidden dimension $d$, number of parameters $N$ of LLMs. D and E refer to dense and MoE models respectively.}
}
\label{tab:llm_size}
\begin{tabular}{@{}lccc@{}}
\toprule
\textbf{Model}   & $d$& $N$ ($10^9$) & $N/d$ ($10^6$)\\ 
\midrule
Mistral~7B (D)~\cite{jiang2023mistral7b} & $4096$& $7.24$ & $1.77$ \\
Mixtral~8x7B (E)~\cite{jiang2024mixtralexperts} & $4096$ & $46.7$ & $11.4$ \\
Qwen2.5~32B (D)~\cite{qwen2025qwen25technicalreport} & $5120$ & $32.8$ & $6.41$ \\
DeepSeek~V3 (E)~\cite{deepseekai2025deepseekv3technicalreport} & $7168$ & $685$ & $95.6$ \\
Llama~3~405B (D)~\cite{dubey2024llama3herdmodels} & $16384$ & $406$& $24.8$ \\
\bottomrule
\end{tabular}
\end{table}

These infrastructure variations drastically affect communication characteristics. To analyze their impact independent of specific hardware or model configurations, we normalize the communication time components ($T_{lat}$ and $T_{bw}$) by the max per-microbatch forward computation time per stage ($T_F$), see Section~\ref{sec:ppdependency}). This yields dimensionless ratios, $T_{lat} / T_F$ and $T_{bw} / T_F$, which capture the relative cost of communication.

\subsection{Cross-DC Parallel Dimension Selection}\label{sec:crossdc_pp_or_dp}
\begin{figure*}[!ht]
\centering
\includegraphics[width=1.0\textwidth]{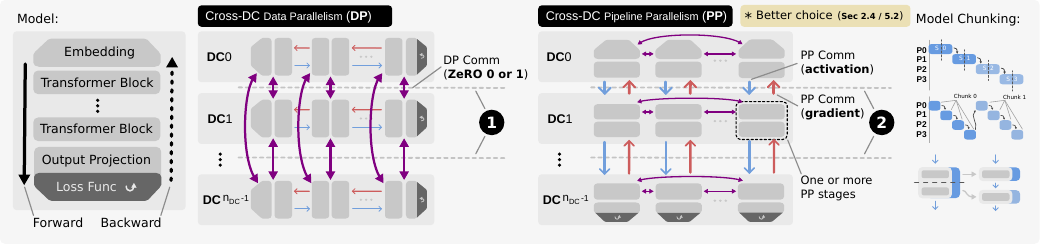}
\caption{
Typical LLM architecture (left). \textbf{Cross-DC DP:} Each DC maintains independent model copies. \includegraphics[scale=1.1,trim=0 1 0 0]{figures/icons/Circle1.pdf} Collective operations (Allgather and Reduce-Scatter, or Allreduce) synchronize gradients and update parameters (ZeRO stage 0 or 1). \textbf{Cross-DC PP:} The model is partitioned among DCs at layer boundaries. Each DC holds one or more pipeline \textit{stages}. DP communication happens internally within each DC to synchronize gradients of stages each holds. \includegraphics[scale=1.1,trim=0 1 0 0]{figures/icons/Circle2.pdf} Inter-DC communication employs point-to-point send/receive operations for exchanging activations and gradients. Stages can further split into finer \textit{chunks} to enhance scheduling efficiency (Section~\ref{sec:ppdependency}).
}
\label{fig:parallelism}
\end{figure*}
\green{Hybrid parallelism exhibits structured communication patterns, with over $99\%$ of GPU pairs having no direct traffic~\cite{wang2024railonlylowcosthighperformancenetwork}. The choice of parallelism strategy across DCs significantly influences training efficiency. Figure~\ref{fig:parallelism} compares two viable options identified in Section~\ref{sec:parallel_strategies}: cross-DC PP and cross-DC DP. Cross-DC PP communication volume is characterized by $sd*n_{DP}$ while cross-DC DP communication volume is characterized by model parameters $N$. The key hyperparameters of some LLMs are shown in Table~\ref{tab:llm_size}. We analyzed both cross-DC PP and cross-DC DP for Llama~3~405B in Section~\ref{sec:analysis_ppvsdp}, demonstrating that cross-DC PP is generally the better choice. The increased popularity of MoE models further shifts the preference towards cross-DC PP~\footnote{Llama~3~405B is trained with $s=8192$, $n_{DP}=128$, yielding $N/sdn_{DP}\approx23.6$. DeepSeek-V3 is trained with $s=4096$, $n_{DP}=128$, yielding $N/sdn_{DP}\approx182$. Notice that Llama~3~405B is wider in $d$ and therefore can be viewed as the basis of a MoE model much larger than DeepSeek-V3.}, since experts in MoE models introduce extra DP communication volume compared to dense models with similar width $d$.}

\section{Pipeline Model}

\subsection{Computation Blocks}

PP partitions model chunks across $n_{PP}$ stages, processing input microbatches in sequence. During the backward pass, gradients propagate in reverse, from the last model chunk back to the initial one. Periods when devices remain idle while awaiting required data are termed pipeline bubbles. To evaluate pipeline efficiency, we define the bubble ratio as the fraction of idle time over total time per device.

We denote the \forward{Forward} computation for each chunk as the \forward{F} block, and the \backward{Backward} computation as the \backward{B} block. Each \backward{B} block can be further split into an input data gradient computation block (\dgrad{DGrad}, or \dgrad{D}) and a weight gradient computation block (\wgrad{WGrad}, or \wgrad{W})~\cite{qi2023zerobubblepipelineparallelism}. An illustration of this decomposition is provided in Appendix~\ref{sec:wgradsplit}. This finer granularity facilitates the construction of more efficient pipeline schedules with reduced bubble ratios.

\subsection{System Parameters}\label{sec:sys_param}
Although our focus is on training Transformer-based~\cite{vaswani2017attention} LLMs in cross-DC environments, our approach applies to large models structured as a sequence of layers. Relevant system parameters include memory usage, per-chunk computation time, and inter-stage communication latency and bandwidth delays. Memory consumption encompasses static elements (parameters, gradients, optimizer states) and dynamic allocations (activations cached during \forward{F} blocks and released after \dgrad{D} and \wgrad{W}).

\subsection{Pipeline Schedules}\label{sec:ppdependency}

\begin{figure}[!h]
\centering
\includegraphics[width=1.0\linewidth]{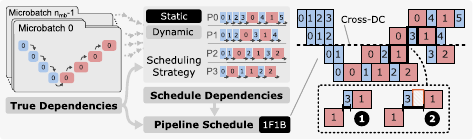}
\caption{
Construction of a 1F1B static schedule. True dependencies guide the creation of schedule dependencies. The resulting acyclic dependency graph governs execution and is used for runtime estimation (Section~\ref{sec:perf_model}).
Highlighted parts (right) show the timing of activation/gradient arrival either \includegraphics[scale=1.1,trim=0 1 0 0]{figures/icons/Circle1.pdf} enables immediate scheduling or \includegraphics[scale=1.1,trim=0 1 0 0]{figures/icons/Circle2.pdf} causes delays (bubbles).  %
}
\label{fig:ppweaving}
\end{figure}

\begin{figure}[!h]
\centering
\includegraphics[width=1.0\linewidth]{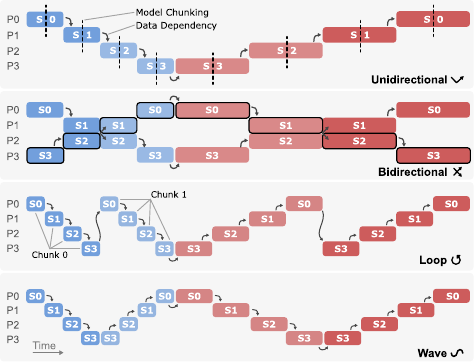}
\caption{
Traversal patterns for a single microbatch (two microbatches shown in the Bidirectional pattern). Loop and Wave patterns leverage model chunking to refine granularity. 
}
\label{fig:pptravesal}
\end{figure}

Pipeline schedules are represented as acyclic dependency graphs, with vertices as pipeline blocks and edges representing two types of dependencies: true dependencies (data dependencies within each microbatch) and schedule dependencies (execution order within each pipeline stage). The construction of a 1F1B schedule is  illustrated in Figure~\ref{fig:ppweaving}.

\textbf{True dependencies} reflect actual data flow across blocks (forward for activations, backward for gradients). Figure~\ref{fig:pptravesal} shows key traversal (data flow) patterns: Unidirectional (UD), Bidirectional (BD), Loop, and Wave.

\textbf{Schedule dependencies} define the execution order of pipeline blocks within each stage. These dependencies are determined using either a static or dynamic scheduling strategy. Static strategies predetermine block placements (usually hand-optimized), while dynamic strategies adapt based on system parameters. CrossPipe schedules (Section~\ref{sec:cp_schedules}) fall under dynamic scheduling.

\subsection{Problems of Static Scheduling}\label{sec:bubble_stride_and_sync}

\brown{In the presence of communication delays, we identify two major limitations that reduce the efficiency of static schedules in cross-DC PP: static execution order (scheduling-level) and static communication arrangement (implementation-level).}

\begin{figure}[!h]
\centering
\includegraphics[width=1.0\linewidth]{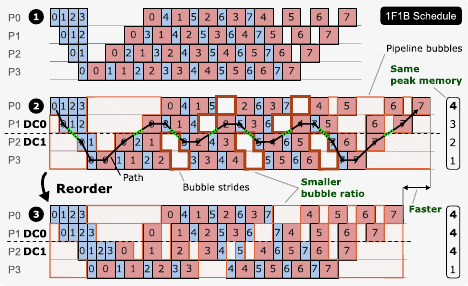}
\caption{
\includegraphics[scale=1.1,trim=0 1 0 0]{figures/icons/Circle1.pdf} Original 1F1B schedule. \includegraphics[scale=1.1,trim=0 1 0 0]{figures/icons/Circle2.pdf} 1F1B schedule with Cross-DC PP communication which leads to \textit{bubble strides}. A \textit{Path} is depicted (\includegraphics[scale=1.8,trim=0 0 0 0]{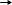}) including cross-DC boundary crossings (\includegraphics[scale=2.5,trim=0 0 0 0]{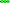}).  \includegraphics[scale=1.1,trim=0 1 0 0]{figures/icons/Circle3.pdf} The schedule after reordering is more efficient while maintaining the same peak memory. More microbatches or memory budget can help to further reduce runtime.
}
\label{fig:reordering}
\end{figure}

\brown{\textbf{Static Execution Order}: Static schedules are optimized under the assumption of negligible communication cost, as in single-DC settings. When directly applied to cross-DC training, they fail to adapt to higher communication delays, resulting in pipeline inefficiencies visualized as \textit{bubble strides}  (illustrated in \includegraphics[scale=1.1,trim=0 1 0 0]{figures/icons/Circle2.pdf} Figure~\ref{fig:reordering} and Appendix~\ref{sec:more_crossdc_pp_examples}). }
\textbf{Schedule} \includegraphics[scale=1.1,trim=0 1 0 0]{figures/icons/Circle2.pdf} in Figure~\ref{fig:reordering} depicts a critical path in a 1F1B schedule across 2 DCs involving 8 cross-DC PP communications. Since the path consists solely of true and schedule dependencies (Section~\ref{sec:ppdependency}), its length imposes a lower bound on overall runtime. For a 1F1B schedule of $n_{mb}$ microbatches, there exists a path containing $O(n_{mb})$ cross-DC communications. As a result, communication delays are amplified proportionally, significantly degrading training throughput.  
A detailed analysis of this amplification effect is presented in Section~\ref{sec:analysis}. CrossPipe addresses this limitation via dynamic scheduling strategies, detailed in Section~\ref{sec:cp_schedules}.
\textbf{Schedule} \includegraphics[scale=1.1,trim=0 1 0 0]{figures/icons/Circle3.pdf} in Figure~\ref{fig:reordering} illustrates that reordering pipeline blocks can improve efficiency while maintaining the same peak activation memory if needed. 
\green{
\textbf{Static Communication Arrangement}: 
Existing frameworks such as Megatron-LM often group pipeline communication operations (e.g., GPU 0 sending to GPU 1 while receiving from GPU 1) for simplicity and hardware efficiency, which introduces implicit synchronization. 
Moreover, even if this grouping is avoided, the two-sided communication pattern introduces synchronization between the sender and receiver in each send/recv operation. Due to variations in stage execution time, the receiver may fail to post the corresponding receive in time, causing the sender to wait.
These delays disrupt stage alignment, which many hand-optimized schedules assume, and propagate bubbles across the pipeline. The interleaved 1F1B schedule~\cite{narayanan2021efficient} overlaps communication with one computation block to mitigate this synchronization cost. However, its static design is only effective under small delays. To address this, CrossPipe decouples scheduling logic from communication orchestration, allowing more fine-grained and adaptive execution, as elaborated in Section~\ref{sec:exeplan}.
}

\subsection{Pipeline Performance Model}\label{sec:perf_model}

\label{sec:pp_perf_model} We develop a performance model to estimate the runtime of a pipeline by leveraging the topological ordering of its dependency graph (Section~\ref{sec:ppdependency}). This model assumes that scheduling and communication orchestration are decoupled, thereby excluding delays caused by synchronization overhead (see Section~\ref{sec:bubble_stride_and_sync}). The start time of each block is determined by the maximum of two values: (1) the completion time of the preceding block on the same stage (\includegraphics[scale=1.1,trim=0 1 0 0]{figures/icons/Circle1.pdf} in Figure~\ref{fig:ppweaving}), and (2) the completion time of the dependent block plus the communication delay (\includegraphics[scale=1.1,trim=0 1 0 0]{figures/icons/Circle2.pdf} in Figure~\ref{fig:ppweaving}). The communication delay consists of a fixed latency component and a bandwidth-related component, which depends on both link bandwidth and current occupancy (Section~\ref{sec:comm_model}).

\section{CrossPipe Schedules}\label{sec:cp_schedules}
\subsection{Optimal Schedule}\label{sec:cpoform}
The optimal pipeline schedule depends on the system parameters and can be framed as a job scheduling problem. Prior work~\cite{qi2023zerobubblepipelineparallelism} formulates this using mixed integer linear programming. Building on this, we elevate communication operations to first-class citizens alongside computation, incorporating both latency and bandwidth delays into the formulation, and generalize this to traversal patterns (Figure~\ref{fig:pptravesal}). This leads us to define the problem as a constraint optimization (CO) task. 
In addition to yielding start and end times for all operations, the solution inherently determines the execution order of communication operations that share the same cross-DC link, thereby handling link contention and scheduling order of cross-DC communications implicitly.

\begin{figure}[!h]
\centering
\includegraphics[width=1.0\linewidth]{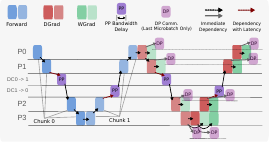}
\caption{\green{The data dependency in a Wave schedule with modeling of both computation and communication. For each model chunk, the DP communication only depends on the \wgrad{W} block of last microbatch. The modeling of vanilla DP is shown in the figure. For DP with ZeRO stage 1, an Allgather block should precede the \forward{F} block of the first microbatch in each model chunk.}
}
\label{fig:wave_dependency}
\end{figure}

\paragraph{Sets and Indices}
\brown{Every \emph{compute} operation is uniquely identified by the
triple $(s, k,t,m)\in\mathcal S \times \mathcal{ K}\times\mathcal T\times\mathcal M$. Each communication operation
$c\in\mathcal C$ transfers the data produced by a compute operation on stage
$\text{src}(c)$ to stage $\text{dst}(c)$.
\begin{itemize}[noitemsep, topsep=0pt,leftmargin=*]
  \item $\mathcal{M}$: Microbatches, indexed by $mb = 1,\dots,n_{mb}$
  \item $\mathcal{S}$: Devices (pipeline stages), indexed by $s = 1,\dots,n_{PP}$
  \item $\mathcal{T}$: Operation types, $\mathcal{T} = \{\forward{F}, \dgrad{D}, \wgrad{W}\}$
  \item $\mathcal{K}$: Model chunks
  \item $\mathcal{P}$: Compute operations (indexed by $(s,k,t,m)$)
  \item $\mathcal{C}$: Communication operations
  \item $\mathcal{O} = \mathcal{P} \cup \mathcal{C}$: All operations
\end{itemize}
}

\paragraph{Inputs} \brown{The input variables are closely related to system parameters in Section~\ref{sec:sys_param}. 
\begin{itemize}[noitemsep, topsep=0pt,leftmargin=*]
  \item $d_o$: Duration of $o \in \mathcal{O}$ (bandwidth time for $c \in \mathcal{C}$, computation time otherwise)
  \item $\ell_c$: Latency delay for $c \in \mathcal{C}$
  \item $m_p$: Net memory change after compute $p \in \mathcal{P}$ completes
  \item $M_s^{\max}$: Memory limit of device $s$
  \item $\text{Pred}(o)$: Immediate predecessor of $o$ in data dependency
\end{itemize}
}

\paragraph{Decision Variables}\brown{
\begin{itemize}[noitemsep, topsep=0pt,leftmargin=*]
  \item $t_o \in \mathbb{R}_{\ge 0}$: Start time of operation $o \in \mathcal{O}$
  \item $x_{o,o'} \in \{0,1\}$: Order for operations sharing a device/link
\end{itemize}
}

\paragraph{Constraints}\brown{
\begin{itemize}[noitemsep, topsep=0pt,leftmargin=*]
\item \textbf{Data Dependencies}  
An example of data dependency of Wave schedules is shown in Figure~\ref{fig:wave_dependency}.
\[
\forall o \in \mathcal{O},\; \forall p \in \text{Pred}(o):\quad
t_o \ge t_p + d_p + 
\begin{cases}
\ell_p & \text{if } p \in \mathcal{C} \\
0 & \text{if } p \in \mathcal{P}
\end{cases}
\]
\item \textbf{Resource Non-overlap}  
For any $o, o' \in \mathcal{O}$ sharing a device or link, overlapping is not allowed:
\[
\begin{aligned}
t_o + d_o &\le t_{o'} + H(1 - x_{o,o'}) \\
t_{o'} + d_{o'} &\le t_o + H x_{o,o'}
\end{aligned}
\]
$H$ is a large constant bounding the scheduling horizon.
\item \textbf{On-device Memory Capacity}  
Let $u_{p,q}=1$ iff compute $p$ completes before $q$ starts. Then for all $s \in \mathcal{S}$ and $q \in \mathcal{P}$ assigned to $s$:
\[
\sum_{\substack{p \in \mathcal{P} \\ \text{device}(p) = s}} m_p\, u_{p,q} \le M_s^{\max}
\]
\item \textbf{Microbatch Order within Stage and Type}  
For any $o,o' \in \mathcal{P}$ on same device $s$ and type $t \in \mathcal{T}$, if microbatch index $mb(o) < mb(o')$, then:
\[
t_o + d_o \le t_{o'}
\]
This constraint reduces the search space.
\end{itemize}
}

\paragraph{Objective}
\brown{We minimize the makespan, defined as the time from the earliest start to the latest finish on the first device:
\[
\min\left( t_{\text{last}(0)} + d_{\text{last}(0)} - t_{\text{first}(0)} \right)
\]
}

\paragraph{DP Overlap Modeling}
\green{DP communication can be modeled as distinct operations triggered after the completion of the \wgrad{W} block for the final microbatch of the current model chunk. In the case of ZeRO stage 1, the corresponding weight Allgather operations are scheduled before the first \forward{F} block of each model chunk. An illustrative example is shown in Figure~\ref{fig:wave_dependency}. The objective is then extended to account for the added communication.}

\paragraph{Solver Scalability}\label{sec:solver_scalability}
We evaluate the runtime performance of both MILP and CO solvers on identical pipeline scheduling problems, as detailed in Appendix~\ref{sec:solver_scalability_comp}. 
It demonstrates the feasibility of using solver-based methods in production environments. However, the search space expands rapidly with the number of pipeline stages, which can lead to prohibitive computation time for extremely large-scale scenarios or dynamically changing system parameters. In such cases, solver-based scheduling becomes less practical, motivating the need for alternative approaches (Section~\ref{sec:greedy_schedule}).

\subsection{Greedy Schedule}\label{sec:greedy_schedule}
To address the scalability limitations of solver-based approaches, we introduce a greedy schedule generation algorithm. This method is designed to rapidly generate near-optimal pipeline schedules while adapting to potentially dynamic system conditions.

\subsubsection{Greedy Sub-block Scheduling}
The greedy algorithm operates using \textit{local} information, only considering scheduled blocks and those ready for scheduling.  To counter the suboptimal decisions typical of greedy methods, we employ block-splitting: each computation block is divided into $n_{sub}$ sub-blocks. This finer granularity enhances the ability to reduce pipeline bubbles with negligible scheduling overhead compared to training iteration time.

\paragraph{Algorithm Inputs} The inputs to the greedy algorithm include:
Per-stage runtimes of \forward{F}, \dgrad{D}, and \wgrad{W} blocks: $T_F$, $T_D$, $T_W$; corresponding memory usage: $M_F$, $M_D$, and $M_W$; per-stage memory limit: $M_L$; communication delay matrices: $\alpha$ and $\beta$, where $\alpha[i,j]$ and $\beta[i,j]$ represent latency and bandwidth inverse of communication from device $i$ to device $j$. The scheduling procedure for the \textit{CrossUDSub} schedule is outlined in Algorithm~\ref{alg:greedy}.

\begin{algorithm}[h]
\footnotesize
\caption{Greedy Generation for \textit{CrossUDSub} Schedule}
\label{alg:greedy}
    \begin{algorithmic}[1]
        \STATE {\bfseries Output:} Per-stage schedule $S_{d}, \forall d\in [n_{PP}]$
        \FOR{$i$ in $[n_{mb}]$} 
            \STATE Add $F_i$ operation for stage 0 to $S_0$'s schedulable operations.
        \ENDFOR
        \WHILE[\hfill Scheduling Loop (Sec~\ref{sec:mainloop})]{True} 
            \STATE $cur \leftarrow \texttt{next\_stage\_to\_schedule()}$
            \IF{no stage is schedulable}
                \STATE\textbf{break}
            \ENDIF
            \STATE $p_{cur} \leftarrow $ schedulable operation of highest priority on stage $cur$
            \STATE Schedule  next sub-block of $p_{cur}$
            \IF{$p_{cur}.type$ = \dgrad{D} \AND no remaining sub-blocks of $p_{cur}$}
                \STATE Add \wgrad{W} operation to schedulable operations of current stage
            \ENDIF
            \STATE Let $p_{next}$ be the operation dependent on $p_{cur}$
            \IF{$p_{next}$ exists \AND  no remaining sub-blocks of $p_{cur}$} 
                \STATE $T_{lat}\leftarrow \alpha[cur, next]$
                \STATE $T_{bw} \leftarrow \beta[cur, next] * \text{Msg\_Size}$
                \STATE $E_{bw} \leftarrow \texttt{bw\_model}(p_{cur}.T_{end}, cur, next, T_{bw}$)
                \STATE $p_{next}.T_{avail} \leftarrow  E_{bw} + T_{lat}$
                \STATE Add $p_{next}$ to $S_{next}$'s schedulable operations.
            \ENDIF
        \ENDWHILE
    \end{algorithmic}
\end{algorithm}

\subsubsection{Scheduling Loop} \label{sec:mainloop}
The scheduling loop consists of three core steps:

\paragraph{Stage Selection}
The \texttt{next\_stage\_to\_schedule} method identifies the stage with the earliest schedulable time, which is defined as the maximum between the end time of the last scheduled operation and the earliest available time of schedulable operations.

\paragraph{Operation Selection} 
The scheduler selects operations available at or after the end of the last scheduled operation on the chosen stage. 
When multiple options exist, it applies a heuristic priority across three phases: 
\begin{itemize}[noitemsep, topsep=0pt, leftmargin=*]
\item \textbf{Warm-up phase}: prioritizes \forward{F} blocks
\item \textbf{Steady phase}: interleaves \forward{F} and \dgrad{D} full blocks
\item \textbf{Tear-down phase}: prioritizes \dgrad{D} over \wgrad{W} blocks
\end{itemize}
When memory constraints prevent scheduling \forward{F} or \dgrad{D} blocks, a \wgrad{W} sub-block is scheduled.

\paragraph{Operation Scheduling}
 The selected operation is scheduled on its stage. If it is the last sub-block of a \dgrad{D} block, the corresponding \wgrad{W} is added to the schedulable operations of the stage. The dependent blocks are then made schedulable on the receiving stage with the earliest start time calculated using the communication model.

\subsubsection{Bandwidth Occupancy Model}\label{sec:bandwidth_model}

To model bandwidth contention, we use a simple range-based bandwidth occupancy model. 
Communication is assumed to begin immediately upon completion of the relevant computation block. The \texttt{BW\_model($T_{ready}, src, dst, T_{bw}$)} function  determines the earliest available transmission window of length $T_{bw}$, starting at or after $T_{ready}$, and returns its end time.

\subsubsection{Performance Characteristics} 
\green{While greedy algorithms do not guarantee global optimality, our approach demonstrates strong empirical performance. As detailed in Section~\ref{sec:analysis}\&\ref{sec:evaluation}, the CrossUDSub schedule achieves:
\begin{itemize}[noitemsep, topsep=0pt,leftmargin=*]
\item Equivalent performance to ZB-H1~\cite{qi2023zerobubblepipelineparallelism} under negligible communication delays and same memory constraints.
\item Faster than static schedules under non-negligible communication delays by filling sub-block size bubbles.
\item Further improvements when memory constraints are relaxed, allowing greater scheduling flexibility.
\end{itemize}}

\subsubsection{Time Complexity}
The main loop executes $3n_{mb}n_{sub}n_{PP}$ iterations, with each iteration scheduling one sub-block. Identifying the next stage and highest priority operation incurs $O(n_{mb}n_{sub}\log(n_{mb}n_{sub}))$ cost. Hence, the overall complexity is $O(n_{mb}^2n_{sub}^2n_{PP}\log(n_{mb}n_{sub}))$. In practice, since the number of schedulable operations per stage remains small, the runtime approximates $O(c \cdot n_{mb}n_{sub}n_{PP})$ for a small constant $c$.

\section{Analysis}\label{sec:analysis}
In this section, we use simulation experiments based on the performance model described in Section~\ref{sec:perf_model} to investigate two key questions: (1) How do different pipeline schedules respond to latency and bandwidth delays? (2) Between cross-DC PP and cross-DC DP, which is more efficient for training in cross-DC settings?

\subsection{Schedule Efficiency}

\begin{table}[h]
\centering
\setlength{\tabcolsep}{1.5pt}
\footnotesize
\sf
\begin{tabular}{@{}lcccccc@{}}
\toprule
\textbf{Schedule} & \textbf{Type} & \textbf{WGrad}  & \textbf{Bubble Ratio} & \textbf{Memory} & \textbf{DP Overlap}\\
\midrule
1F1B~\cite{narayanan2019pipedream} & UD  & Combined & High & Medium & Medium\\
IV1F1B~\cite{narayanan2021efficient} & Loop  & Combined & Medium & Medium+ & Medium+\\
ZBH1~\cite{qi2023zerobubblepipelineparallelism} & UD  & Split & Medium & Medium & Low\\
ZBV~\cite{qi2023zerobubblepipelineparallelism} & Wave & Split & Low & Medium & Low\\
\bottomrule
\end{tabular}
\rm
\caption{Static pipeline schedules used in the analysis. UD and BD stands for unidirectional and bidirectional. IV1F1B is the abbreviation for the interleaved 1F1B schedule.}
\label{tab:ppschedules}
\end{table}

\begin{figure}[!h]
\centering
\includegraphics[width=1.0\linewidth]{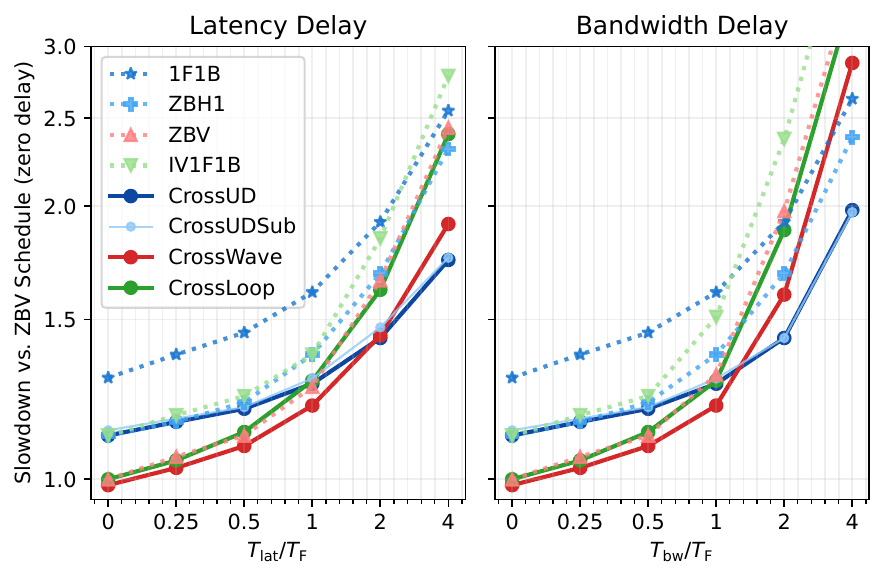}
\caption{
Impact of latency and bandwidth delay on runtime across different pipeline schedules. Static (\includegraphics[scale=0.6,trim=0 -2.3 0 0]{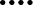}) and dynamic (\includegraphics[scale=0.6,trim=0 -2.3 0 0]{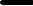}) schedules are compared. Setup: $4$ stages, $8$ microbatches ($2$ DCs, $2$ stages per DC). \textit{Cross}-prefixed schedules are generated by the CO solver (Section~\ref{sec:cpoform}). Slowdown is measured relative to the ZBV schedule at zero delay.
}
\label{fig:pre_delay_sensitivity}
\end{figure}

We compare various pipeline schedules under increasing communication delay in a cross-DC PP setting. Bidirectional (BD) schedules are excluded as they involve both PP and DP cross-DC communication. The main static schedules that we focus on are summarized in Table~\ref{fig:pre_delay_sensitivity}. We use $n_{PP}=4, n_{mb}=8$, and simulate 2 DCs with 2 stages each.  Dynamic schedules are generated with the same memory limits as their static counterparts (e.g., CrossUD mirrors 1F1B). Delay sensitivity is measured as slowdown relative to the ZBV schedule under no communication delay. Delay is varied using $T_{lat}/T_F$ (latency delay) and $T_{bw}/T_F$ (bandwidth delay), where $T_F$ is the per-stage forward computation time. 
Key observations from Figure~\ref{fig:pre_delay_sensitivity} include:
\begin{itemize}[noitemsep, topsep=0pt, leftmargin=*]
\item WGrad-split schedules consistently outperform unified-backward ones due to finer scheduling granularity.
\item Wave schedules are more efficient in low-delay settings, while UD schedules become superior as delays grow.
\item Loop schedules show the highest sensitivity to delays, due to more frequent cross-DC communication (6 per microbatch, compared to 4 for Wave and 2 for UD).
\item The greedy CrossUDSub schedule matches the solver-based CrossUD in most delay regimes, highlighting its efficacy as a lightweight alternative.
\item When delays are small, latency and bandwidth contribute equally to runtime. However, once bandwidth delay exceeds the forward time per chunk, it induces additional pipeline bubbles from queuing (Section~\ref{sec:comm_model}).
\end{itemize}

\subsection{Cross-DC PP vs. Cross-DC DP}\label{sec:analysis_ppvsdp}

We simulate iteration times for cross-DC PP and cross-DC DP approaches using the Llama~3~405B model~\cite{dubey2024llama3herdmodels} under various latency and bandwidth conditions (detailed in Appendix~\ref{sec:sim_crossdc_pp_dp}). 

\begin{figure}[!h]
\centering
\includegraphics[width=1.0\linewidth]{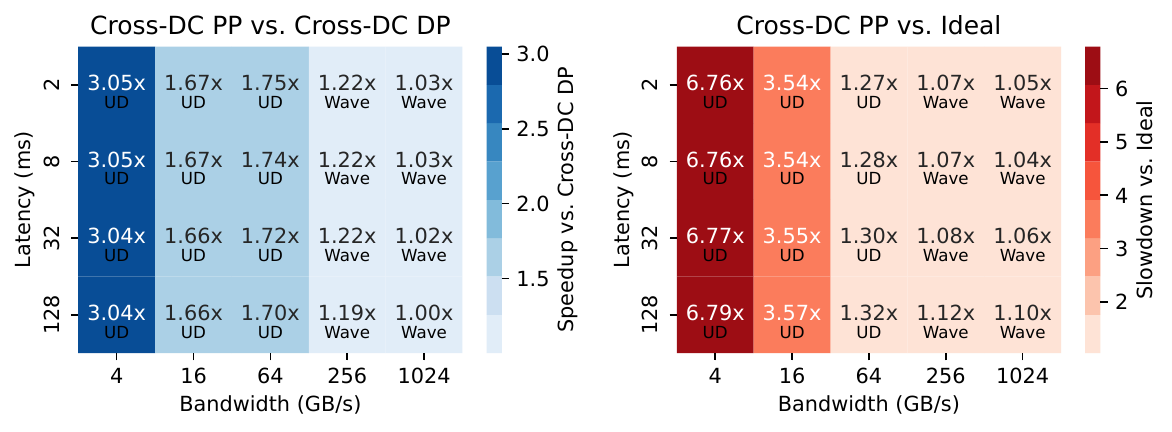}
\caption{
Simulation results comparing cross-DC PP and DP for Llama~3~405B training across two DCs. Left: Speedup of cross-DC PP over cross-DC DP. Right: Slowdown of cross-DC PP compared to an ideal single-DC setup. Labels indicate exact values and optimal schedule types per configuration.
}
\label{fig:cross_dc_dp_or_pp}
\end{figure}

Results in Figure~\ref{fig:cross_dc_dp_or_pp} show that latency (ranging from 4–128~ms) has little impact on runtime in this scenario, as the per-stage forward time ($T_F\approx 109$ms) keeps the delay ratio low. However, bandwidth significantly affects performance. Cross-DC PP outperforms cross-DC DP by up to 3.05x when the cross-DC link bandwidth is limited to 4~GB/s. This gap narrows as bandwidth increases, becoming negligible beyond 1024~GB/s. Compared to the ideal single-DC case, cross-DC PP sees only a 1.3× slowdown at 64~GB/s. These results suggest that for large models with long per-stage computation time ($T_F$), bandwidth is the primary bottleneck in cross-DC communication. Under such conditions, cross-DC PP offers superior efficiency relative to DP, particularly when network resources are constrained.

\section{CrossPipe Implementation} \label{sec:implementation}

Schedules generated by CrossPipe can adapt to configuration changes, including PP size, hybrid parallelism setups, and system parameters. In contrast, static PP modules in existing frameworks support only a limited, hard-coded range of schedules, making them difficult to adapt to and extend. Our implementation addresses these limitations through the CrossPipe module, which integrates seamlessly with the existing training framework. We use Megatron-LM as our base framework. The CrossPipe module is primarily implemented in Python, with components in C++ to enable latency and bandwidth injection (Section~\ref{sec:latbwinject}) for emulating cross-DC network conditions on a homogeneous cluster.

\begin{figure}[!h]
\centering
\includegraphics[width=1.0\linewidth]{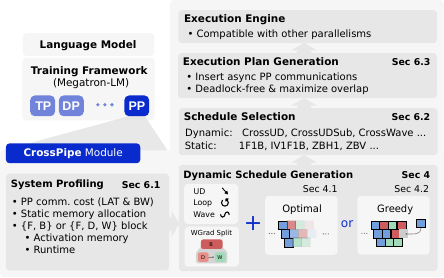}
\caption{
Components of the CrossPipe module.
}
\label{fig:implementation}
\end{figure}

An overview of our implementation is shown in 
Figure~\ref{fig:implementation}
. The module begins by collecting system parameters via lightweight benchmarks (Section~\ref{sec:sysprof}). It then generates dynamic pipeline schedules (defining the order and timing of computation blocks) using either the constraint optimization solver (Section~\ref{sec:cpoform}) or the greedy algorithm (Section~\ref{sec:greedy_schedule}). A schedule with the best simulation performance is selected (Section~\ref{sec:schedsel}). Next, CrossPipe lowers it to a concrete execution plan by inserting and optimizing communication operations (Section~\ref{sec:exeplan}).  This decouples high-level scheduling logic from low-level execution, enabling dynamic, fine-grained control.
The selected schedule can also be hot-swapped during training if better options are found.

\subsection{System Profiling}\label{sec:sysprof}

CrossPipe collects critical metrics in a single iteration using lightweight profiling. These include runtime and memory usage of \forward{F}, \dgrad{D}, and \wgrad{W} blocks, as well as communication delay parameters ($\alpha$, $\beta$). We follow the model partitioning strategy of Llama~3~\cite{dubey2024llama3herdmodels}, treating embedding and output layers as transformer layers to ensure load balance across stages.

\subsection{Schedule Selection}\label{sec:schedsel}
CrossPipe selects the schedule with the best estimated performance  and supports hot-switching to adapt to changes during training. 

\begin{itemize}[noitemsep, topsep=0pt,leftmargin=*]
\item\textbf{Static schedules} are well-suited for single-DC training.
\item \textbf{Dynamic schedules} are more suitable in cross-DC settings with high or varying communication costs. They also adapt better to  available memory. Under rich memory budgets, dynamic schedules may increase the number of in-flight \forward{F} blocks to improve efficiency.
\end{itemize}

\subsection{Execution Plan}\label{sec:exeplan}
In this step, CrossPipe converts the selected pipeline schedule into an execution plan by inserting non-blocking communication operations. This plan is executed by the CrossPipe engine integrated into the training framework.

\paragraph{Communication Orchestration} We use NCCL as the communication backend to leverage high-bandwidth intra-node interconnects and reduce inter-node data movement overheads. To decouple point-to-point communications in PP, we dedicate four GPU streams for each direction and role (\{\text{Send}, \text{Recv}\} × \{\text{Next}, \text{Prev}\}), avoiding interference and deadlocks. In both directions, Recv operations are reordered to align with the corresponding Sends, avoiding NCCL deadlocks. NCCL implements a rendezvous protocol for point-to-point communication, requiring both the sender and the receiver to synchronize before the transfer begins. To maximize communication overlap, we post Recv operations ahead of their corresponding Sends based on profiling estimates. This delay-aware arrangement improves overlap and is applied to both static and dynamic schedules (evaluated in Section~\ref{sec:evaluation}).

\subsection{Latency and Bandwidth Injection}\label{sec:latbwinject}

We extend the PyTorch \texttt{ProcessGroupNCCL} C++ backend to inject latency and bandwidth delays in specific Send/Recv operations. 
This allows us to emulate various cross-DC network conditions (as described in Section~\ref{sec:infra}) within a single cluster. Latency is injected on the receiver side of cross-DC communication, while bandwidth is throttled by running spinning kernels on the communication streams of both sender and receiver. Implementation details are provided in Appendix~\ref{sec:appendixlatbwinject}, and validation results in Appendix~\ref{sec:delayinjvalid}.

\section{Evaluation}\label{sec:evaluation}
\begin{figure*}[!htbp]
    \centering
    \begin{subfigure}[t]{0.49\textwidth}
        \centering
        \includegraphics[width=\textwidth]{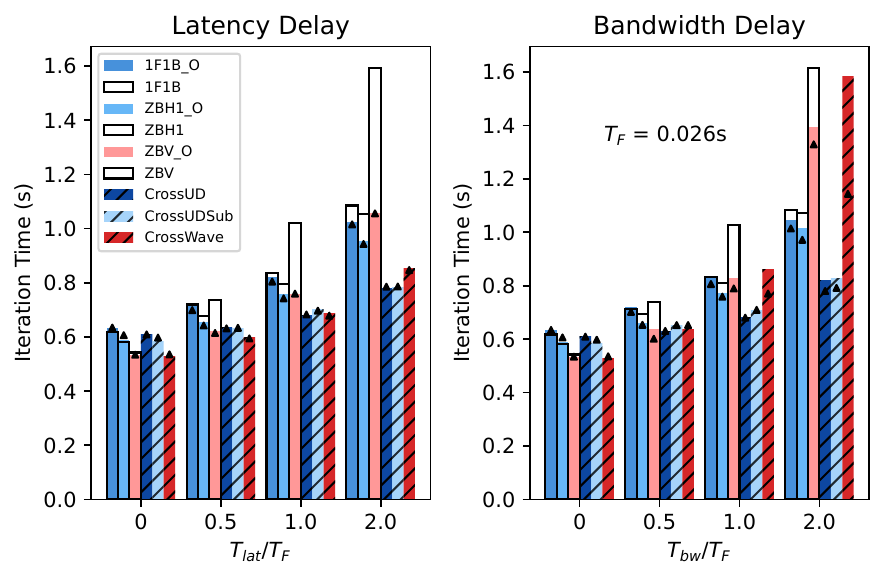}
        \caption{Model M8}
        \label{fig:subfig1}
    \end{subfigure}
    \hfill
    \begin{subfigure}[t]{0.49\textwidth}
        \centering
        \includegraphics[width=\textwidth]{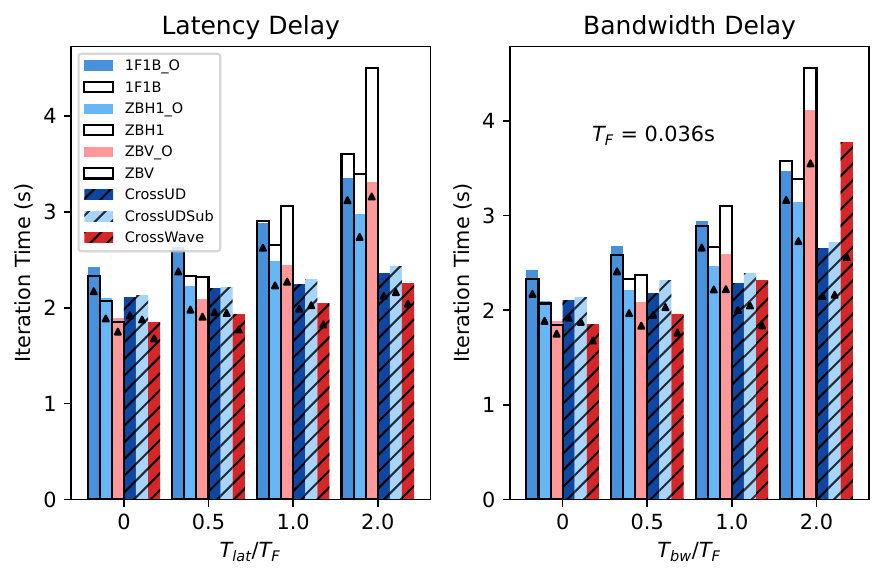}
        \caption{Model M70}
        \label{fig:subfig2}
    \end{subfigure}
    \caption{Evaluation of static and CrossPipe schedules under various emulated latency and bandwidth delay ratios. The runs of static schedules with communication arrangement optimization (Section~\ref{sec:exeplan}) is marked with the suffix \_O. The runtime prediction from the performance model (Section~\ref{sec:perf_model}) is represented as (\includegraphics[scale=0.4,trim=0 0 0 0]{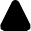}). Schedules are compared using the iteration time in seconds (lower is better). Dynamic schedules (Cross*) use the same peak memory budget as their corresponding static base schedule (e.g., CrossUD matches 1F1B).
    }
    \label{fig:exp0_lat_bw}
\end{figure*}
We conducted comprehensive evaluations on the Alps supercomputer to validate CrossPipe's performance and scalability. Each compute node is equipped with four GH200 Grace Hopper Superchips~\cite{nvidia_grace_hopper_2024}. Each GH200 features 96~GB HBM3 memory integrated with the Hopper GPU die and 120~GB LPDDRX5 memory connected to the Grace CPU. The chips utilize a fully-connected topology with six NVLink 4.0 links between each GH200 pair, providing 200~Gb/s bandwidth per link per direction. Network connectivity is provided by HPE Cray Cassini-1 200~Gb/s NICs in a Dragonfly~\cite{kim2008technology} topology using HPE Cray Slingshot-11~\cite{shafie2022high, de2020depth} interconnect.

\begin{table}
\small
\setlength{\tabcolsep}{4pt} 
\footnotesize
\sf
\begin{tabular}{lccccc}
\toprule
\textbf{Name} & \textbf{Hidden Dim.} & \textbf{Int. Dim.} & \textbf{Att. Heads} & \textbf{KV Heads} & \textbf{Layers}\\
\midrule
M8  & 4096 & 14336 & 32 & 8  & 30+2\\
M70  & 8192 & 28672 & 64 & 8 & 62+2\\
\bottomrule
\end{tabular}
\rm
\caption{Hyperparameters of models used in the evaluation. Example: M70 is a model with Transformers layers of the same size as the ones in the Llama~3~70B model. The number of layers is reported as number of transformer layers + embedding \& output layers.}
\label{tab:llama_configs}
\end{table}

We use LLMs built up from the Llama-style Transformer layers, the hyperparameters of which are listed in Table~\ref{tab:llama_configs}. Each model configuration follows the naming convention \texttt{M\{model\_size\}}, where the model\_size indicates the size of Transformer layers it contains. For example, M70 uses the Transformer block of the same size as the one in the Llama~3~70B model. The models are then constructed by replicating and stacking identical Transformer layers, along with the vocabulary embedding layer and the output layer. We evaluate the following schedules: 1F1B, IV1F1B, ZBH1, ZBV, CrossUD, CrossUDSub, and CrossWave. The reason to exclude bidirectional (BD) and Loop schedules is explained in Section~\ref{sec:analysis}. By default, we set the microbatch size $b=1$ and sequence length $s=4096$. \green{In these configurations, the message size for pipeline communication (activations/gradients) is approximately $32n_{DP}$ MB for M8 and $64n_{DP}$ MB for M70. }FlashAttention~\cite{shah2024flashattention3fastaccurateattention} is enabled for higher throughput and less peak memory consumption. For each schedule, we measure 64 iterations and report the minimum value to minimize the network noise effects~\cite{networknoise}.

\subsection{Impact of Latency and Bandwidth Delay}\label{sec:exp0_lat_bw_delay}

We evaluate the performance of schedules on the cluster with various emulated latency delay $T_{lat}$ or bandwidth delay $T_{bw}$ for each PP communication crossing the DC boundary, using the injection mechanism from Section~\ref{sec:latbwinject}. We conduct the experiments on both M8 and M70 models in a two-DC setting. The parallelism configurations are: 
\begin{itemize}[noitemsep, topsep=0pt,leftmargin=*]
    \item \textbf{M8:} $n_{TP}=2$, $n_{PP}=4$, $n_{DP}=1$, $GBS=2n_{PP}n_{DP}=8$.
    \item \textbf{M70:} $n_{TP}=4$, $n_{PP}=8$, $n_{DP}=1$, $GBS=2n_{PP}n_{DP}=16$.
\end{itemize}
We vary delay ratios  $T_{lat}/T_F,T_{bw}/T_F \in \{0.0, 0.5, 1.0, 2.0\}$. $T_F$ is defined as the maximum runtime of the per-microbatch forward computation among stages. The results are shown in Figure~\ref{fig:exp0_lat_bw}. For the static schedules, we vary two settings: with or without the delay-aware communication orchestration  (Section~\ref{sec:exeplan}). Runs with this optimization are marked with the suffix \_O. This optimization reduces the impact of delayed receivers, aligning closely with the assumptions of our performance model (Section~\ref{sec:pp_perf_model}). This model accurately predicts the runtime of tested schedules in most configurations. Overall, the CrossPipe schedules show superior performance compared to static schedules, with a reduction in runtime of up to $33.6\%$ (vs. original) or $21.9\%$ (vs. optimized), achieved by the M70 model at $T_{bw}/T_F=2$.

\green{To ground these delay ratios in realistic scenarios, we take the M70 model as an example. We assume a practical DP size of $n_{DP}=16$, with intra-DC DP communication fully overlapped. Under this setting, the message size per PP communication is calculated by $bsdn_{DP} * 2 = 1$~GB. Given a forward time of $T_F=0.038$~s (Figure~\ref{fig:exp0_lat_bw}.b), the resulting injected latency delays range from 19~ms to 76~ms, and simulated bandwidth from 105~Gbps to 421~Gbps.}

\subsection{Further Bubble Reduction}\label{sec:exp1_extra_gbs_mem}

Section~\ref{sec:exp0_lat_bw_delay} shows that block reordering helps to reduce PP runtime when the memory budget and global batch size (GBS) are strictly constrained. However, substantial bubble ratios persist under high communication costs. This observation necessitates examination of trade-offs among runtime, GBS and memory budget. Also, we take layer-wise activation recomputation~\cite{chen2016trainingdeepnetssublinear} into consideration. Since the GBS varies among the settings, we use runtime per microbatch to compare schedules. We conduct the experiments on the M70 model in a 2-DC setting, with $n_{TP}=4, n_{PP}=8, n_{DP}=1$, under delay combinations: $(T_{lat}/T_F,T_{bw}/T_F) \in \{(0,0), (0.25, 2), (2, 0.25), (2,2)\}$, across three configurations:
\begin{itemize}[noitemsep, topsep=0pt,leftmargin=*]
    \item \textbf{Case 1:} $GBS=2n_{PP}n_{DP}=16$, activation memory budget $1.0\times$ (same as 1F1B), no recomputation.
    \item \textbf{Case 2:} $GBS=32$, activation memory budget $1.0\times$, layer-wise recomputation.
    \item \textbf{Case 3:} $GBS=32$, activation memory budget $2.0\times$, no recomputation.
\end{itemize}

\begin{table}[htbp]
\small
\footnotesize
\sf
\addtolength{\tabcolsep}{-1pt}    
\begin{tabular}{@{}ccc|cccccc@{}}
\toprule
\multirow{2}{*}{$\frac{T_{lat}}{T_F}$} & \multirow{2}{*}{$\frac{T_{bw}}{T_F}$} & \multirow{2}{*}{\textbf{Case}} & \multicolumn{3}{c}{\textbf{Static}} & \multicolumn{3}{c}{\textbf{Dynamic (This Work)}} \\
\cmidrule(lr){4-6} \cmidrule(lr){7-9}
& & & \textbf{1F1B} & \textbf{ZBH1} & \textbf{ZBV} & \textbf{UDSub} & \textbf{UD} & \textbf{Wave} \\ 
\midrule
\multirow{3}{*}{0} & \multirow{3}{*}{0} & 1 & 0.151 & 0.133 & \textbf{0.118} & 0.137 & 0.137 & 0.119 \\
 &  & 2 & 0.174 & 0.168 & 0.161 & - & 0.165 & \textbf{0.157} \\
 &  & 3 & - & - & - & 0.121 & 0.118 & \textbf{0.108} \\ \midrule
\multirow{3}{*}{0.25} & \multirow{3}{*}{0.25} & 1 & 0.168 & 0.15 & 0.148 & 0.149 & 0.142 & \textbf{0.127} \\
 &  & 2 & 0.193 & 0.187 & 0.177 & - & 0.17 & \textbf{0.159} \\
 &  & 3 & - & - & - & 0.123 & 0.123 & \textbf{0.112} \\ \midrule
\multirow{3}{*}{0.25} & \multirow{3}{*}{2} & 1 & 0.241 & 0.23 & 0.315 & 0.181 & \textbf{0.177} & 0.25 \\
 &  & 2 & 0.262 & 0.259 & 0.33 & - & \textbf{0.185} & 0.274 \\
 &  & 3 & - & - & - & \textbf{0.144} & 0.15 & 0.235 \\ \midrule
\multirow{3}{*}{2} & \multirow{3}{*}{0.25} & 1 & 0.242 & 0.229 & 0.314 & 0.16 & 0.153 & \textbf{0.148} \\
 &  & 2 & 0.262 & 0.258 & 0.329 & - & 0.173 & \textbf{0.163} \\
 &  & 3 & - & - & - & 0.127 & 0.124 & \textbf{0.115} \\ \midrule
\multirow{3}{*}{2} & \multirow{3}{*}{2} & 1 & 0.321 & 0.309 & 0.473 & 0.198 & \textbf{0.196} & 0.256 \\
 &  & 2 & 0.333 & 0.331 & 0.476 & - & \textbf{0.196} & 0.291 \\
 &  & 3 & - & - & - & \textbf{0.145} & 0.147 & 0.245 \\ \bottomrule
\end{tabular}
\rm
\caption{M70 model, 2-DC training. Runtime per microbatch of cross-DC PP solutions under various configurations (case 1-3) and communication delay. The best result of each configuration is shown in bold (lower is better). CrossUDSub, CrossUD and CrossWave schedules are abbreviated as UDSub, UD and Wave, respectively.}
\label{tab:extra_gbs_mem}
\end{table}

The result is shown in Table~\ref{tab:extra_gbs_mem}. Static schedules cannot leverage extra memory budget to further reduce pipeline bubbles ('-' in case 3). Without delay, where $(T_{lat}/T_F,T_{bw}/T_F) = (0,0)$, the efficiency of the CrossPipe schedules is comparable to manually optimized static schedules. Increasing GBS amortizes bubbles in the warm-up and tear-down phases by extending the length of the steady phase which contains fewer bubbles. Under high latency (e.g.,$(2,0.25)$), increasing GBS and memory budget helps the CrossWave schedule to achieve pipeline efficiency (0.115~s per microbatch), matching the no delay case (0.118~s per microbatch). When the bandwidth delay dominates (e.g., $(2,0.25)$ or $(2,2)$), increasing both GBS and memory budget improves schedule efficiency by up to $1.33\times$ (0.196~s to 0.147~s per microbatch for CrossUD). In general, the bandwidth delay is harder to mitigate than the latency delay under the same settings of GBS and memory budget. Layer-wise recomputation generally does not improve the runtime of dynamic schedules, as the recomputation during the backward pass negates its low memory benefits.

\subsection{Scale to More DCs}
We extend our analysis to 4 homogeneous interconnected DCs with uniform cross-DC link characteristics, using the same setup as the previous section (M70, $n_{TP}=4$, $n_{PP}=8$, $n_{DP}=1$, now with $2$ stages per DC). 
Table~\ref{tab:4dc} confirms previous findings: CrossPipe is competitive without delay and outperforms static schedules in cross-DC scenarios. CrossWave excels CrossUD(Sub) at low delays but suffers under higher delays, especially under high bandwidth delays. With extra memory budget and GBS (Case 3), CrossUD schedule achieves 0.178~s per microbatch, only 22.8\% slower than the corresponding 2-DC scenario at $(T_{lat}/T_F,T_{bw}/T_F) = (2,2)$. The larger bubble size in 4-DC training makes recomputation more effective here. Layer-wise recomputation with increased GBS (Case 2) is comparable to or outperforms the baseline without recomputation (Case 1) in most delay settings.

\begin{table}[htbp]
\small
\footnotesize
\sf
\addtolength{\tabcolsep}{-1pt}    
\begin{tabular}{@{}ccc|cccccc@{}}
\toprule
\multirow{2}{*}{$\frac{T_{lat}}{T_F}$} & \multirow{2}{*}{$\frac{T_{bw}}{T_F}$} & \multirow{2}{*}{\textbf{Case}} & \multicolumn{3}{c}{\textbf{Static}} & \multicolumn{3}{c}{\textbf{Dynamic (This Work)}} \\
\cmidrule(lr){4-6} \cmidrule(lr){7-9}
& & & \textbf{1F1B} & \textbf{ZBH1} & \textbf{ZBV} & \textbf{UDSub} & \textbf{UD} & \textbf{Wave} \\ 
\midrule
\multirow{3}{*}{0} & \multirow{3}{*}{0} & 1 & 0.149 & 0.133 & \textbf{0.119} & 0.138 & 0.138 & 0.122 \\
 &  & 2 & 0.173 & 0.168 & 0.16 & - & 0.167 & \textbf{0.157} \\
 &  & 3 & - & - & - & 0.123 & 0.119 & \textbf{0.115} \\ \midrule
\multirow{3}{*}{0.25} & \multirow{3}{*}{0.25} & 1 & 0.177 & 0.158 & 0.161 & 0.155 & 0.148 & \textbf{0.141} \\
 &  & 2 & 0.198 & 0.19 & 0.181 & - & 0.173 & \textbf{0.163} \\
 &  & 3 & - & - & - & 0.126 & 0.123 & \textbf{0.115} \\ \midrule
\multirow{3}{*}{0.25} & \multirow{3}{*}{2} & 1 & 0.269 & 0.249 & 0.339 & \textbf{0.216} & 0.217 & 0.286 \\
 &  & 2 & 0.274 & 0.269 & 0.331 & - & \textbf{0.198} & 0.29 \\
 &  & 3 & - & - & - & \textbf{0.158} & 0.162 & 0.262 \\ \midrule
\multirow{3}{*}{2} & \multirow{3}{*}{0.25} & 1 & 0.268 & 0.248 & 0.337 & 0.2 & \textbf{0.197} & 0.214 \\
 &  & 2 & 0.274 & 0.269 & 0.33 & - & 0.184 & \textbf{0.177} \\
 &  & 3 & - & - & - & \textbf{0.138} & \textbf{0.138} & 0.139 \\ \midrule
\multirow{3}{*}{2} & \multirow{3}{*}{2} & 1 & 0.359 & 0.338 & 0.512 & \textbf{0.268} & 0.271 & 0.339 \\
 &  & 2 & 0.349 & 0.346 & 0.479 & - & \textbf{0.213} & 0.295 \\
 &  & 3 & - & - & - & \textbf{0.178} & \textbf{0.178} & 0.264 \\ 
 \bottomrule
\end{tabular}
\caption{M70 model, 4-DC training. Runtime per microbatch of each PP schedule is listed. The rest of the configurations remain the same as Table~\ref{tab:extra_gbs_mem}.}
\label{tab:4dc}
\end{table}

\subsection{Trade-off of PP and DP}\label{sec:eval_ppvsdp}

\begin{figure}[]
    \centering
    \begin{subfigure}[t]{0.49\columnwidth}
        \centering\captionsetup{justification=raggedleft}
        \includegraphics[width=\textwidth]{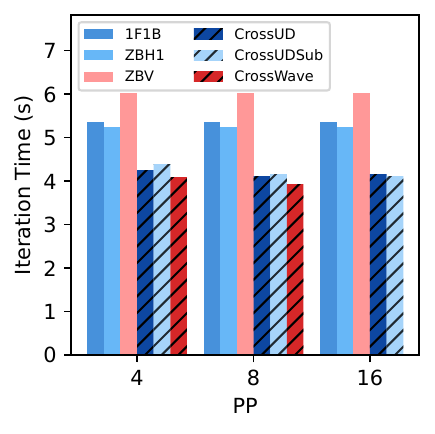}
        \caption{
        $T_{lat}/T_F = 1, T_{bw}/T_F = 0.25$; Static Mem. ($n_{PP}=4$): 25\%
        }
        \label{fig:subfig1}
    \end{subfigure}
    \hfill
    \begin{subfigure}[t]{0.49\columnwidth}
        \centering\captionsetup{justification=raggedleft}
        \includegraphics[width=\textwidth]{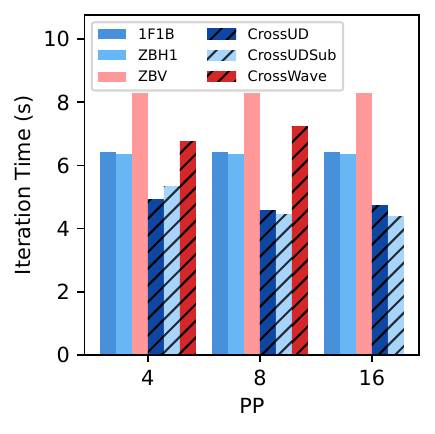}
        \caption{
        $T_{lat}/T_F = 0.125, T_{bw}/T_F = 2$; Static Mem. ($n_{PP}=4$): 25\%
        }
        \label{fig:subfig2}
    \end{subfigure}
    
    \vspace{1em}
    
    \begin{subfigure}[t]{0.49\columnwidth}
        \centering\captionsetup{justification=raggedleft}
        \includegraphics[width=\textwidth]{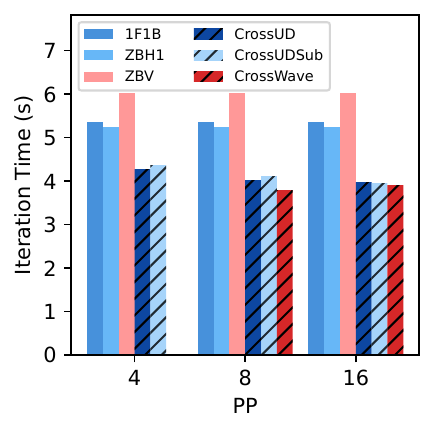}
        \caption{
        $T_{lat}/T_F = 1, T_{bw}/T_F = 0.25$; Static Mem. ($n_{PP}=4$): 50\%
        }
        \label{fig:subfig3}
    \end{subfigure}
    \hfill
    \begin{subfigure}[t]{0.49\columnwidth}
        \centering\captionsetup{justification=raggedleft}
        \includegraphics[width=\textwidth]{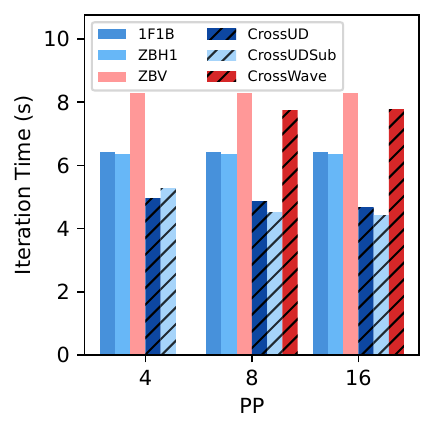}
        \caption{
        $T_{lat}/T_F = 0.125, T_{bw}/T_F = 2$; Static Mem. ($n_{PP}=4$): 50\%
        }
        \label{fig:subfig4}
    \end{subfigure}
    \caption{Trade-off between PP and DP ($n_{PP}\times n_{DP}$ is fixed) in a 2-DC training. Using M70 with 16 nodes ($n_{TP}=4$ fixed) and a fixed GBS. Each subplot shows iteration time vs. $n_{PP}$. Labels indicate delay ratios ($T_{lat}/T_F$, $T_{bw}/T_F$) and static memory usage. Memory in percentage shows the consumption of static memory at $n_{PP}=4$ w.r.t. device limit.
    }
    \label{fig:pp_dp_tradeoff}
\end{figure}

We analyze the choice of large PP size vs. large DP size in a 2-DC homogeneous training scenario. With a fixed number of compute nodes per DC and setting $n_{TP}$ to the number of GPUs within each compute node, the product $n_{PP}\times n_{DP}$ equals the total node count of 2 DCs. Given the number of microbatches $n_{mb}=\epsilon n_{PP}$, the global batch size $GBS=\epsilon n_{PP}n_{DP}$ depends solely on the ratio $\epsilon$. Then the GBS remains the same with various combinations of PP and DP. 

We evaluated the trade-off on the M70 model across three PP configurations ($n_{PP}\in\{4,8,16\}$) with four combinations of communication delay and memory settings. 
For communication delay $(T_{lat}/T_F,T_{bw}/T_F)$, we test high latency, low bandwidth $(1,0.25)$ and low latency, high bandwidth $(0.125,2)$ settings, with a static memory budget of 25\% and 50\%. This percentage is defined as the allocation of static memory (parameters, gradients, and optimizer states) at $n_{PP}=4$ w.r.t. the device memory limit. \brown{Larger $n_{PP}$ reduces the static memory per stage and the activation memory size per microbatch, so that it increases the activation memory budget and allows for more in-flight forward blocks. Also with larger $n_{PP}$, the communication volume is lower due to smaller DP, but at a higher frequency. While the absolute latency remains constant when increasing $n_{PP}$, the latency delay ratio $T_{lat}/T_F$ increases due to reduced $T_F$ as $n_{PP}$ grows (less work per stage). $\epsilon=4$ to enable more in-flight blocks in each schedule to benefit from available memory. Since the GBS (total workload) and the number of compute nodes (total GPUs) are fixed, we use runtime to compare different configurations. As shown in Figure~\ref{fig:pp_dp_tradeoff}, these factors balance out. The schedule efficiency remains largely invariant to PP/DP configurations across scenarios. This balance likely stems from the flexibility of CrossPipe to fully utilize the available memory with sufficient GBS.}

\section{Discussion}
\subsection{Heterogeneous DCs}
In this work, we primarily evaluated homogeneous GPUs and nodes across DCs due to cluster constraints.  However, our approach can be extended to heterogeneous environments. If compute resources (e.g., GPU, network) differ significantly between DCs, our generated schedules may be suboptimal because faster nodes will finish computation earlier and create  bubbles in the pipeline, and slower nodes may run out of memory since their device memory is usually more limited. \brown{A practical solution is to maintain homogeneity within each pipeline stage, which aligns with current DC practices where nodes within the same cabinet tend to be identical. The compute nodes with faster GPUs can be assigned more layers to balance the computation time across stages. Our formulation (Section~\ref{sec:cpoform}) and greedy algorithm (Section~\ref{sec:greedy_schedule}) already support stage-specific parameters, naturally supporting this adjustment.}

\subsection{Network Dynamics and Fault Tolerance}
\green{
Real-world cross-DC networks exhibit variability and are prone to failures. CrossPipe can employ several strategies to enhance robustness:
\begin{itemize}[noitemsep, topsep=0pt,leftmargin=*]
\item Short-Term Variations (seconds or less): Section~\ref{sec:exp1_extra_gbs_mem} shows that CrossPipe schedules can trade system resources (e.g., device memory, by allowing more in-flight microbatches) for efficiency. Conservative (higher) latency and/or (lower) bandwidth estimations can be used to generate schedules that tolerate small spikes in communication delay.
\item Longer-Term Variations (minutes or more): Network conditions can shift due to traffic or routing. CrossPipe's flexible execution engine (Section~\ref{sec:implementation}) supports hot-switching of pipeline schedules. The system with CrossPipe enabled can periodically re-profile network conditions (Section~\ref{sec:sysprof}) and generate new, tailored greedy schedules (Section~\ref{sec:greedy_schedule}) to adapt without interrupting training.
\item Packet Loss/Link Errors: Transient network errors such as packet drops can be handled by mechanisms like Forward Error Correction (FEC)~\cite{6772729} or Selective Repeat~\cite{anagnostou2003performance}, leading to spikes in communication delay as mentioned above.
\item Node Failures: The failure of a compute node requires higher-level mechanisms beyond pipeline scheduling. Efficient checkpointing, such as asynchronous methods~\cite{mohan2021checkfreq} and in-memory approaches~\cite{10.1145/3600006.3613145}, is necessary to save training state efficiently and recover from the last checkpoint with minimal progress loss.
\end{itemize}
}

\section{Related Works}
\textbf{Pipeline Parallelism (PP)}: 1F1B in PipeDream~\cite{narayanan2019pipedream}, GPipe~\cite{huang2019gpipe}, DAPPLE~\cite{fan2020dapplepipelineddataparallel}, interleaved 1F1B~\cite{narayanan2021efficient}, bidirectional pipeline from Chimera~\cite{li2021chimera}, BPipe~\cite{pmlr-v202-kim23l} and MPress~\cite{10071077} for memory balancing, BFSPP for more DP communication overlap~\cite{lamypoirier2023breadthfirstpipelineparallelism}, Hanayo~\cite{liu2023hanayo} for wave-like schedules, ooo backprop~\cite{oh2022out} and zero-bubble PP~\cite{qi2023zerobubblepipelineparallelism} for weight gradient splitting, AdaPipe~\cite{adapipe} for co-optimizing layer distribution and recomputation, DHelix~\cite{dhelix} for microbatch co-execution to overlap communication, DistMM~\cite{distmm} for multimodal model training. Sequence-level pipeline parallelism~\cite{pmlr-v139-li21y, sun2024seq1f1befficientsequencelevelpipeline}.

\noindent\textbf{Training on restricted networks with PP}: Varuna~\cite{athlur2021varunascalablelowcosttraining} explores training on spot VMs with commodity networking. Bamboo~\cite{thorpe2022bamboomakingpreemptibleinstances} studies resilient training on preemptible instances. Oobleck~\cite{Jang_2023} improves training resilience via pipeline templates. SWARM~\cite{ryabinin2023swarmparallelismtraininglarge} proposes reliable training via temporary randomized pipelines. ~\cite{yuan2023decentralizedtrainingfoundationmodels} studies device assignment in hybrid parallel training with geo-distributed nodes. CocktailSGD~\cite{cocktailsgd} combines multiple compression techniques to train models efficiently on low-bandwidth networks. FusionLLM~\cite{tang2024fusionllmdecentralizedllmtraining} accelerates decentralized training via activation and gradient compression. DiLoCo~\cite{douillard2024dilocodistributedlowcommunicationtraining} explores robust asynchronous training on poorly connected machines.

\section{Conclusion}
In this work, we first introduced a validated pipeline performance model that explicitly accounts for latency and bandwidth delays in cross-datacenter links. Using this model, we demonstrated that pipeline parallelism is often the superior approach for the parallelism dimension spanning across datacenters, especially under constrained network conditions. Next, we leveraged the model to develop optimal and near-optimal algorithms for generating pipeline schedules that minimize cross-datacenter communication delays while adhering to memory constraints. Finally, we integrated these methods into a flexible execution engine featuring a two-layer abstraction (block scheduling and communication arrangement) that works seamlessly with existing training systems, such as Megatron-LM.

Our evaluation shows that CrossPipe effectively overcomes the challenges of cross-datacenter training. It reduces the training time by up to 33.6\% compared to traditional pipeline schedules in a cross-DC setup, all while maintaining the same memory constraints. When memory constraints are relaxed, CrossPipe in a cross-DC setup is able to achieve a similar training time as a static ZBV schedule in a single-DC setup where there is almost no communication delay. CrossPipe thus offers improved scalability and resource utilization, making large-scale distributed training more feasible and efficient.

\section*{Acknowledgments}
We are grateful to our shepherd and the anonymous reviewers for their insightful comments and constructive feedback.
We thank the CSCS team for providing access to the Ault and Alps machines, as well as for their outstanding technical support.
We are grateful to Siyuan Shen and Mikhail Khalilov for their valuable advice, and to Timo Schneider for his assistance with infrastructure at SPCL.
We also acknowledge the Polish high-performance computing infrastructure PLGrid (HPC Center: ACK Cyfronet AGH) for providing computational resources and support.

\clearpage

\bibliographystyle{plain}
\bibliography{example_paper}

\appendix
\clearpage
\section{Runtime Definition}
In cross-DC training, due to the relatively high communication cost, the stage 0 usually finishes last among all the stages. So we adopt the most used definition of pipeline runtime as the following: the duration from the start of the first forward block on PP stage 0 till the completion of the last block on any PP rank (or the final DP communication on any rank). This definition aligns with frameworks that apply additional synchronization for global gradient norm computation (for clipping) and numerical anomaly detection (NANs/INFs) in mixed-precision training.

\section{Weight Gradient Separation} \label{sec:wgradsplit}

\begin{figure}[!h]
\centering
\includegraphics[width=0.7\linewidth]{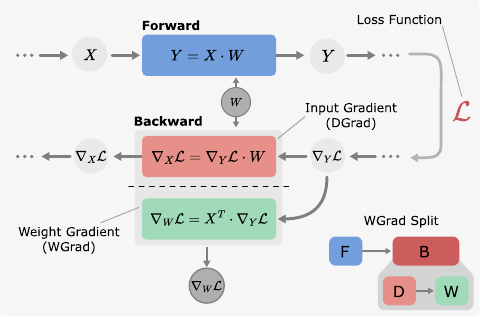}
\caption{
Separating input gradient computation (\dgrad{DGrad}, or \dgrad{D}) and weight gradient computation (\wgrad{WGrad}, or \wgrad{W}) in a linear layer.
}
\label{fig:wgradsplit}
\end{figure}

\noindent Figure~\ref{fig:wgradsplit} illustrates how each \backward{B} block can be further divided into two parts: input data gradient computation (\dgrad{DGrad}, or \dgrad{D}) and weight gradient computation (\wgrad{WGrad}, or \wgrad{W})~\cite{qi2023zerobubblepipelineparallelism}.

\begin{figure}[!h]
\centering
\includegraphics[width=1.0\linewidth]{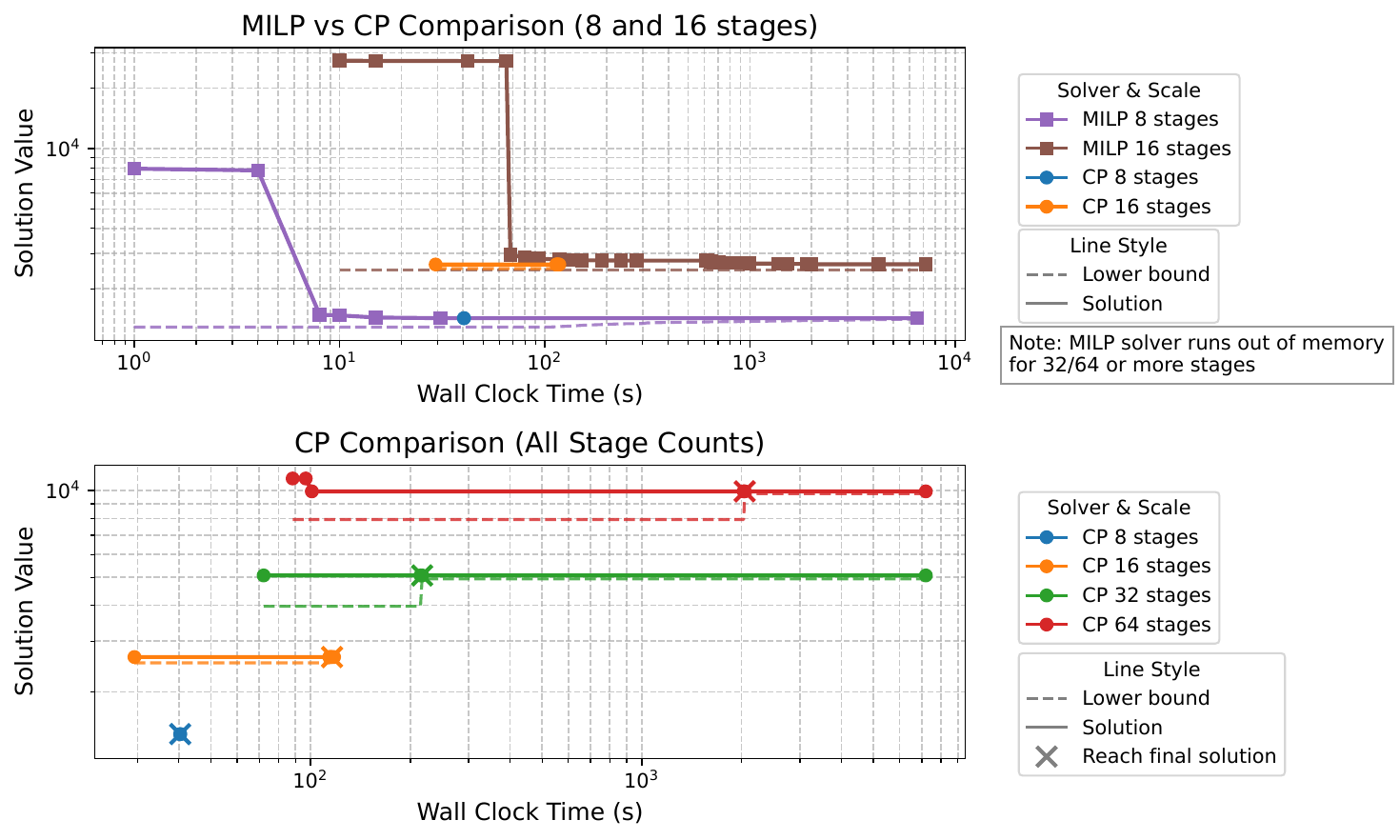}
\caption{
Scalability of MILP and CP solvers finding optimal pipeline schedules. Results show time-to-solution for varying pipeline stage counts ($n_{PP}$) with a fixed microbatch factor ($n_{mb}=2n_{PP}$). Markers indicate the moments when the solver discovers an improved solution.
}
\label{fig:solverscale}
\end{figure}

\section{Bubble Strides in Schedules}\label{sec:more_crossdc_pp_examples}

\begin{figure*}[!ht]
\centering
\includegraphics[width=1.0\textwidth]{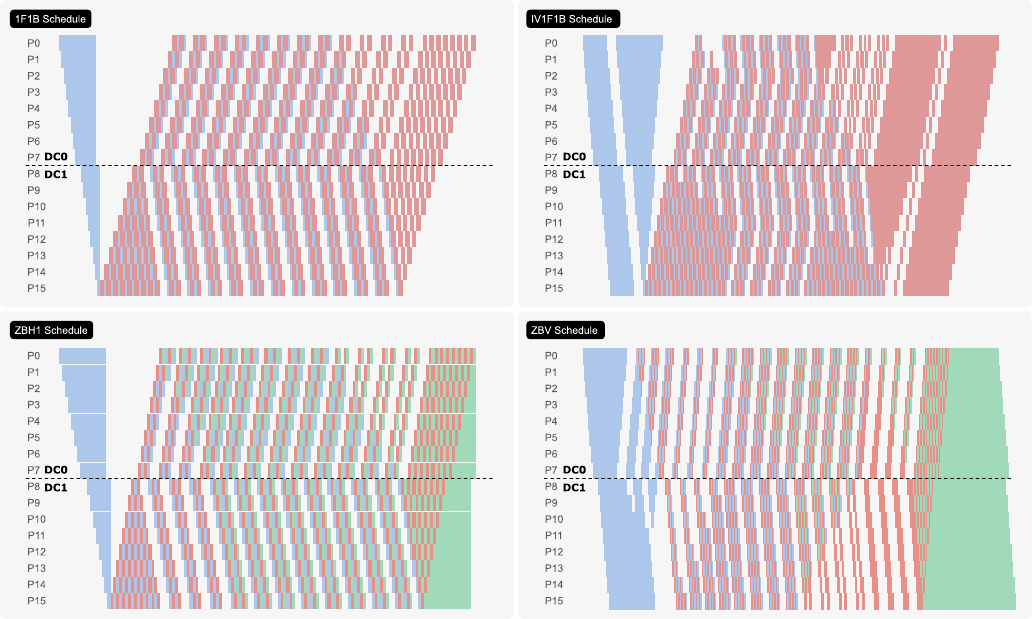}
\caption{
Illustration of bubble strides in various static pipeline schedules (1F1B, IV1F1B, ZBH1, and ZBV). Setup: 16 stages, 2DCs (8 stages per DC), $n_{mb}=32$.
}
\label{fig:appendix_bubble_strides}
\end{figure*}

Figure~\ref{fig:appendix_bubble_strides} further illustrates the existence of \textit{bubble strides}. We show schedules with 16 stages and 2 DCs for the static schedules that we compared in our work (1F1B, IV1F1B, ZBH1, and ZBV). The latency delay on the cross-DC link is set to $1.5 \cdot T_{F}$. Despite this relatively small latency (compared to the size of the pipeline blocks), the delay accumulates throughout the schedule rather than being absorbed or mitigated. This accumulation underscores the need for adaptive scheduling.

\section{Optimal Scheduling}
\subsection{Scalability of MILP and CO Solvers}\label{sec:solver_scalability_comp}

We compare the scalability of both MILP and CO solvers, Gurobi (12.0.0) and CPLEX (22.1.1) respectively, both among the best solvers in their fields. The MILP formulation is from \cite{qi2023zerobubblepipelineparallelism} and the CO formulation from Section~\ref{sec:cpoform}. We conduct the experiments on a machine with AMD EPYC 7742 @ 2.25GHz CPU (128 physical cores), 256GB RAM. Each solver is configured to 256 worker threads, with a time limit of 7200 seconds. We use Gurobi Optimizer with the \texttt{NodefileStart=0.5} parameter to handle potential memory limitations.The solver is early-terminated if the relative gap between the objective value of the best integer solution found so far (ObjVal) and the best objective bound (ObjBound) is below $1\%$, defined as $\frac{|ObjBound-ObjVal|}{|ObjVal|}\leq 0.01$. The time to solution of both solvers are shown in Figure~\ref{fig:solverscale}. The CO solver scales better than the MILP solver for pipeline schedule generation, ideally because of its specialized optimization w.r.t. job scheduling problems. The MILP solver runs out of memory even for runs of 32/64 stages and converges slower than CO solver. On the other hand, CO solver shows tractable performance when problem size scales up, and is able to find a good feasible solution in a reasonable amount of time while spending most of the time searching for a tighter bound.

\section{Comparing Cross-DC PP and Cross-DC DP}\label{sec:sim_crossdc_pp_dp}
We use a 2 DC setup for the Llama 3 405B model training. The training configuration is taken from Llama 3~\cite{dubey2024llama3herdmodels}: $n_{TP}=8$, $n_{PP}=16$, $n_{DP}=64$, $seq\_len=8192$, $GBS=2n_{PP}n_{DP}$. 

We estimate the computation time of each layer by the following equation: $$T_{layer}=\frac{C_{layer}}{P_{GPU}\times n_{TP}}$$ where $C_{layer}$ is the FLOPs count of a transformer layer in the 405B model, $P_{GPU}$ is the practical BF16 performance of GPU (500 TFLOPs per second is used in the analysis, considering the TP communication overhead). Assuming the GPUs are equally distributed in 2 DCs.  In cross-DC PP, we apply solver-based schedules  (CrossUD or CrossWave, whichever is better for the given delay). In cross-DC DP, we use the ZBV schedule and assume that the extra cross-DC communication is not overlapped. For Allreduce (or Reduce-Scatter + Allgather) DP communications, we apply the bandwidth-optimal Ring-algorithm. The cross-DC DP communication cost can be estimated as $2\times (\alpha + 2\times\frac{N}{2}\times\beta) = 2\alpha + 2N\beta$ which accounts for two rounds of communication between two DCs and half of the parameters/gradients are sent/received. $N$ is the number of model parameters, the extra factor $2$ in the bandwidth term comes from the size of the BF16 datatype (2 bytes/param).

\section{Latency and Bandwidth Delay Injection}

\subsection{Implementation Details}\label{sec:appendixlatbwinject}

\begin{figure}[!h]
\centering
\includegraphics[width=1.0\linewidth]{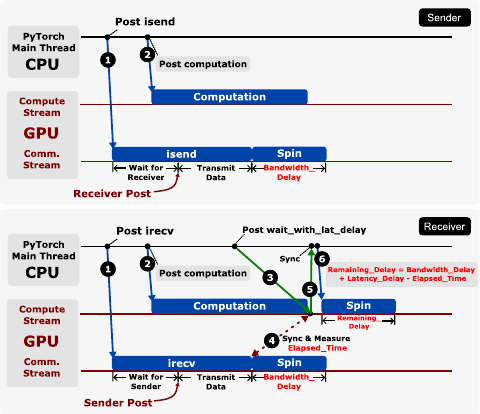}
\caption{
Mechanism for latency and bandwidth delay injection. Top: sender side. Bottom: receiver side. 
}
\label{fig:latbwinject}
\end{figure}

Figure~\ref{fig:latbwinject} illustrates our mechanism for injecting latency and bandwidth delay, used to emulate cross-DC network behavior in a controlled setting, without modifying core NCCL behavior or requiring network hardware manipulation.
GPU kernel execution, including that of NCCL communication kernels, is asynchronous to the host CPU, making precise delay injection non-trivial. 
\begin{itemize}[noitemsep, topsep=0pt,leftmargin=*]
\item \textbf{Bandwidth Delay}: To simulate limited bandwidth, we inject additional delay into the communication path by occupying the communication stream after each NCCL send/receive. Specifically, a spin kernel is posted to the same stream immediately after the NCCL call, both on the sender and receiver sides. 
The spin kernel duration is computed to match the target delay. This effectively stalls further communication or computation that shares the stream, emulating a fully utilized link.

\item \textbf{Latency Delay}: 
We extend PyTorch NCCL backend by adding a new method, \texttt{handle.wait\_with\_lat\_delay()},  which is invoked on the receiver side during each blocking wait on NCCL communication. It is similar to the original \texttt{wait()} method but adds a controlled amount of host-side spinning to delay the launch of subsequent computation kernels.
\end{itemize}

The injection process contains the following steps: \includegraphics[scale=1.1,trim=0 1 0 0]{figures/icons/Circle1.pdf} Sender and receiver post matching asynchronous point-to-point operations.\includegraphics[scale=1.1,trim=0 1 0 0]{figures/icons/Circle2.pdf} Optional compute kernels may be posted to overlap with communication. \includegraphics[scale=1.1,trim=0 1 0 0]{figures/icons/Circle3.pdf} The receiver calls \texttt{wait\_with\_lat\_delay} to synchronize the computation stream with the communication stream. \includegraphics[scale=1.1,trim=0 1 0 0]{figures/icons/Circle4.pdf} A CUDA event measures the elapsed time from the completion of the communication kernel to the current time on the computation stream. \includegraphics[scale=1.1,trim=0 1 0 0]{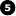} The host then synchronizes with the computation stream to retrieve the elapsed time and calculates the remaining delay (if any) to inject. \includegraphics[scale=1.1,trim=0 1 0 0]{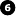} A spinning kernel is posted to the computation stream for the remaining delay. This mechanism ensures that delay is not introduced when communication has already completed, and only partially injected when communication is partially complete (e.g., due to overlap with computation), thereby preserving the correct performance behavior of latency-hiding schedules.

This injection method is specifically designed for the CrossPipe implementation, where communication in four directions is split across four concurrent streams. Injecting delays into collective communications is more complex. While latency delays can be introduced within communication libraries~\cite{shen2024llampassessingnetworklatency}, simulating bandwidth delays may require additional configurations at the network switch level.

In our communication model, we account only for latency and bandwidth delays along the critical path. We assume that the transmission of control messages, such as "ready to send" signals in a rendezvous protocol, is removed from the critical path. This assumption is crucial for maintaining performance, especially under high-latency conditions.

\subsection{Validation}\label{sec:delayinjvalid}

\begin{figure}[!h]
\centering
\includegraphics[width=1.0\linewidth]{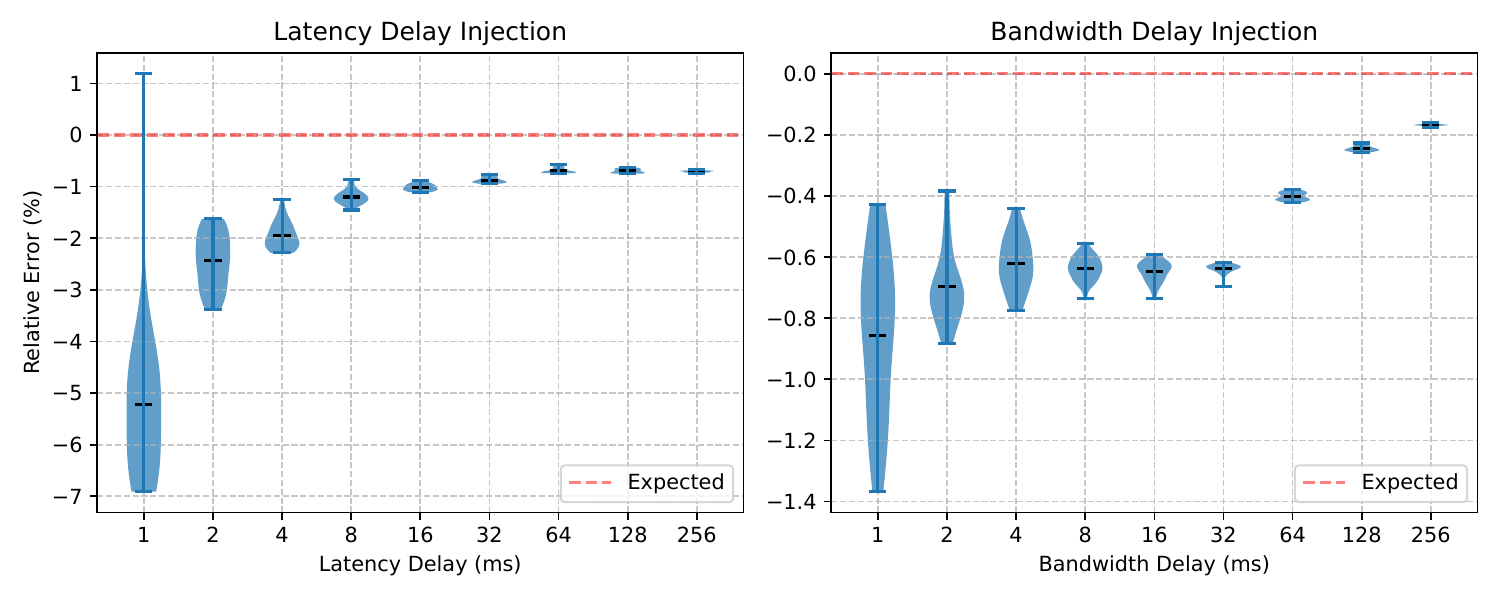}
\caption{
Validation of latency and bandwidth delay injection methods described in Section~\ref{sec:latbwinject}.
}
\label{fig:injection_validation}
\end{figure}

\noindent To validate the accuracy of our delay injection methods (Section~\ref{sec:latbwinject}), we conduct experiments on a single GH200 node using 4 GB messages.  

For latency tests, we use ping-pong communication between two processes: each round consists of a sender transmitting a message, waiting for a reply, and measuring the round-trip time. We divide the round-trip latency by two and compare it against the baseline (no injection). For bandwidth tests, the sender transmits multiple large messages back-to-back, while the receiver posts matching receives. We compute the average time per message and compare it to the baseline.

For each delay setting, we compute the relative error between the observed and expected delays. Specifically, we perform multiple iterations per setting, discard outliers, and report the percentage deviation from the expected delay. These measurements are then visualized using violin plots.

Results in Figure~\ref{fig:injection_validation} demonstrate that our injection methods accurately reflect the communication model described in Section~\ref{sec:comm_model}. Minor deviations primarily stem from host CPU synchronization and slight underestimation of GPU clock rate by \texttt{cudaGetDeviceProperties} in the spinning kernel. The validation shows the injected delay closely matches the target delay, confirming the mechanism's suitability for emulating cross-DC network conditions.

\end{document}